\documentclass[a4paper,11pt]{article}
\usepackage{aaskaiid}
\usepackage{natbib}
\usepackage{url}
\usepackage{comment}
 \usepackage{ulem}
\usepackage{orcidlink}

\newcommand{\blue}[1]{\textcolor{blue}{ #1}}

\usepackage{newunicodechar}
\newunicodechar{ʼ}{'}

\newcommand\HII{\hbox{H$\,\rm \scriptstyle II$}~} 

\title{Exploring the Cosmic Dawn through the 21 cm Forest and High-redshift Radio Sources with the SKA
 }
\ShortTitle{21cm forest with SKA}

\author[1,2]{Junsong Cang\orcidlink{0000-0002-0061-0728}}
\author[3]{Benedetta Ciardi}
\author[4]{Barun Maity\orcidlink{0000-0002-4682-6970}}
\author[5,6]{Atsushi J. Nishizawa\orcidlink{0000-0002-6109-2397}}
\author[7]{Qi Niu\orcidlink{0009-0007-1168-0928}}
\author[8]{Yue Shao\orcidlink{0000-0001-6620-2826}}
\author[9]{Abinash Kumar Shaw\orcidlink{0000-0002-6123-4383}}
\author[10,11,12]{Hayato Shimabukuro\orcidlink{0000-0003-4850-0656}}
\author[13,14,15]{Tomáš Šoltinský\orcidlink{0000-0001-7703-8929}}
\author[7]{Tian-Yang Sun\orcidlink{0009-0002-5109-6420}}
\author[12,16]{Tsutomu T.\ Takeuchi\orcidlink{0000-0001-8416-7673}}
\author[17,18]{Yidong Xu\orcidlink{0000-0003-3224-4125}}
\author[19]{Kohji Yoshikawa\orcidlink{0000-0003-0389-5551}}
\author[17,18]{Bin Yue\orcidlink{0000-0002-7829-1181}}
\author[7,20,21]{Xin Zhang\orcidlink{0000-0002-6029-1933}}

\affiliation[1]{Theoretical and Scientific Data Science, Scuola Internazionale Superiore di Studi Avanzati (SISSA), Via Bonomea 265, 34136 Trieste, Italy}
\affiliation[2]{Scuola Normale Superiore, Piazza dei Cavalieri 7, 56126 Pisa, Italy}
\affiliation[3]{Max Planck Institute for Astrophysics, Karl-Schwarzschild-Str. 1, 85741 Garching, Germany}
\affiliation[4]{Max-Planck-Institut für Astronomie, Königstuhl 17, D-69117 Heidelberg, Germany}
\affiliation[5]{Gifu Shotoku Gakuen University, 1--1 Takakuwanishi, Yanaizucho, Gifu, 501-6122, Japan}
\affiliation[6]{Kobayashi Maskawa Institute, Nagoya University, Furocho, Chikusaku, Nagoya, 461-8602, Japan}
\affiliation[7]{Key Laboratory of Cosmology and Astrophysics (Liaoning), College of Sciences, Northeastern University, Shenyang 110819, China}
\affiliation[8]{Department of Physics, Liaoning Normal University, Dalian 116029, China}
\affiliation[9]{Department of Computer Science, University of Nevada, Las Vegas, NV 89154, USA}
\affiliation[10]{South-Western Institute for Astronomy Research (SWIFAR), Yunnan University, Kunming, Yunnan 650500, People’s Republic of China}
\affiliation[11]{Key Laboratory of Survey Science of Yunnan Province, Yunnan University, Kunming, Yunnan 650500, Peopleʼs Republic of China}
\affiliation[12]{Nagoya University, Graduate School of Science, Division of Particle and Astrophysical Science, Chikusa-Ku, Nagoya, 464-8602, Japan}
\affiliation[13]{INAF–Osservatorio Astronomico di Trieste, Via G.B. Tiepolo, 11, I-34143 Trieste, Italy}
\affiliation[14]{IFPU, Institute for Fundamental Physics of the Universe, Via Beirut 2, I-34151 Trieste, Italy}
\affiliation[15]{INFN, Sezione di Trieste, Via Valerio 2, I-34127 Trieste, Italy}
\affiliation[16]{The Research Center for Statistical Machine Learning, the Institute of Statistical Mathematics, 10--3 Midori-cho, Tachikawa, Tokyo 190--8562, Japan}
\affiliation[17]{National Astronomical Observatories, Chinese Academy of Sciences, Beijing 100101, People's Republic of China}
\affiliation[18]{State Key Laboratory of Radio Astronomy and Technology, Beijing 100101, People's Republic of China}
\affiliation[19]{Center for Computational Sciences, University of Tsukuba, 1--1--1 Tennodai, Tsukuba, Ibaraki 305--8571, Japan}
\affiliation[20]{National Frontiers Science Center for Industrial Intelligence and Systems Optimization, Northeastern University, Shenyang 110819, China}
\affiliation[21]{Key Laboratory of Data Analytics and Optimization for Smart Industry (Ministry of Education), Northeastern University, Shenyang 110819, China}

\emailAdd{shimabukuro@ynu.edu.cn}

\abstract{The 21~cm forest, manifesting as absorption features in the spectra of distant radio sources, is caused by neutral hydrogen in the intervening cosmic structures. It provides a unique opportunity to directly probe the neutral intergalactic medium (IGM) during the epoch of reionization. In particular, this phenomenon offers a distinctive probe of small-scale structures during the epoch of reionization (EoR), and can be used to sensitively constrain the thermal history of the early universe. However, real measurement of the 21~cm forest signals has been hindered by the high sensitivity required to detect the weak individual absorption lines, and the unavailability of high-redshift radio-bright sources at the EoR. In the past decade, significant progress has been made to make the 21~cm forest measurement actually feasible and promising. These include the development of new statistical observables to enhance the sensitivity of measurements with a reasonable observation time, the revision of the expected number counts of background radio-loud quasars in light of recent observations, and the development of deep-learning-based tools to extract physical information more efficiently and accurately. More importantly, methods have been proposed to break the degeneracy between astrophysical effects from the early galaxies and various physical effects that influence the formation of small-scale structures. Therefore, with the advent of the Square Kilometre Array (SKA), it is very promising to utilize the 21~cm forest to probe both the heating effect from early structure formation, or possibly from exotic energy-injection processes, and a range of fundamental physics, such as the dark matter properties, the neutrino mass, the running spectral index, and the relative velocity between dark matter and baryons. In this chapter, we provide a comprehensive overview of recent efforts in the field of 21~cm forest research, and discuss observational strategies and prospects for constraining the first galaxies and the fundamental physics with 21~cm forest observations using the upcoming SKA-Low.
}

\begin{document}
\maketitle

\newcommand{\actaa}{Acta Astron.} 
\newcommand{\araa}{Annu. Rev. Astron. Astrophys.} 
\newcommand{\aar}{Astron. Astrophys. Rev.} 
\newcommand{\ab}{Astrobiol.} 
\newcommand{\aj}{Astron. J.} 
\newcommand{\apj}{Astrophys. J.} 
\newcommand{\apjl}{Astrophys. J. Lett.} 
\newcommand{\apjs}{Astrophys. J. Suppl. Ser.} 
\newcommand{\ao}{Appl. Opt.} 
\newcommand{\apss}{Astrophys. Space Sci.} 
\newcommand{\aap}{Astron. Astrophys.} 
\newcommand{\aapr}{Astron. Astrophys. Rev.} 
\newcommand{\aaps}{Astron. Astrophys. Suppl.} 
\newcommand{\baas}{Bull. Am. Astron. Soc.} 
\newcommand{\caa}{Chinese Astron. Astrophys.} 
\newcommand{\cjaa}{Chinese J. Astron. Astrophys.} 
\newcommand{\cqg}{Class. Quantum Gravity} 
\newcommand{\gal}{Galaxies} 
\newcommand{\gca}{Geochim. Cosmochim. Acta} 
\newcommand{\icarus}{Icarus} 
\newcommand{\jcap}{J. Cosmol. Astropart. Phys.} 
\newcommand{\jgr}{J. Geophys. Res.} 
\newcommand{\jgrp}{J. Geophys. Res.: Planets} 
\newcommand{\jqsrt}{J. Quant. Spectrosc. Radiat. Transf.} 
\newcommand{\memsai}{Mem. Soc. Astron. Italiana} 
\newcommand{\mnras}{Mon. Not. R. Astron. Soc.} 
\newcommand{\nat}{Nature} 
\newcommand{\nastro}{Nat. Astron.} 
\newcommand{\ncomms}{Nat. Commun.} 
\newcommand{\nphys}{Nat. Phys.} 
\newcommand{\na}{New Astron.} 
\newcommand{\nar}{New Astron. Rev.} 
\newcommand{\physrep}{Phys. Rep.} 
\newcommand{\pra}{Phys. Rev. A} 
\newcommand{\prb}{Phys. Rev. B} 
\newcommand{\prc}{Phys. Rev. C} 
\newcommand{\prd}{Phys. Rev. D} 
\newcommand{\pre}{Phys. Rev. E} 
\newcommand{\prl}{Phys. Rev. Lett.} 
\newcommand{\psj}{Planet. Sci. J.} 
\newcommand{\planss}{Planet. Space Sci.} 
\newcommand{\pnas}{Proc. Natl Acad. Sci. USA} 
\newcommand{\procspie}{Proc. SPIE} 
\newcommand{\pasa}{Publ. Astron. Soc. Aust.} 
\newcommand{\pasj}{Publ. Astron. Soc. Jpn} 
\newcommand{\pasp}{Publ. Astron. Soc. Pac.} 
\newcommand{\rmxaa}{Rev. Mexicana Astron. Astrofis.} 
\newcommand{\sci}{Science} 
\newcommand{\sciadv}{Sci. Adv.} 
\newcommand{\solphys}{Sol. Phys.} 
\newcommand{\sovast}{Soviet Ast.} 
\newcommand{\ssr}{Space Sci. Rev.} 
\newcommand{\uni}{Universe} 


\section{Overview of the 21~cm Forest}

The 21~cm forest refers to the absorption features caused by neutral hydrogen (HI) observed against high-redshift (high-\(z\)) radio-loud sources such as quasars. This phenomenon occurs when cold neutral hydrogen in the intergalactic medium (IGM) or minihalos along the line of sight absorbs radiation from background sources, imprinting characteristic absorption lines onto their spectra. It is a unique method to probe the early universe, complementing approaches such as tomography and power spectrum analyses while providing distinct insights\cite[e.g.][]{Carilli_2002, Furlanetto_2002,Furlanetto_2006}.

An instructive analogy is the well-known Lyman-$\alpha$ forest, which is absorption lines from neutral hydrogen seen in quasar spectra\citep[e.g.][]{Weinberg_2003,McQuinn_2016}. Like the 21~cm forest, it probes the IGM along the line of sight, but differs in key ways. The Lyman-$\alpha$ forest appears in the UV/optical and traces warmer, ionized or partially neutral gas, while the 21~cm forest, seen in radio, is sensitive to cold neutral hydrogen. Moreover, whereas the Lyman-$\alpha$ forest is mainly used at the end stage of reionization ($z \lesssim 6$), the 21~cm forest can explore higher redshifts during the initial stage of reionization. 

One of the key quantities that characterize the 21~cm forest is the optical depth \(\tau_{\nu}\), which quantifies the degree of absorption experienced by radio waves due to neutral hydrogen\cite[e.g.][]{Furlanetto_2002,Furlanetto_2006}. The optical depth of a cloud of hydrogen can be expressed as
\begin{equation}
\tau_{\nu} = \frac{3h_p c^3A_{\rm 10}}{32\pi^{3/2}k_{\rm B}}\,
\frac{1}{\nu^2}\int_{-\infty}^{+\infty} 
\frac{n_{\rm HI}(x)}{b(x)T_{\rm S}(x)}\,
\exp{\left[-\frac{(u(\nu)-v(x))^2}{b^2(x)}\right]}\, {\rm d}x
\label{eq:tau21}
\end{equation}

Here, \(c\) is the speed of light, \(h_p\) is Planck's constant, \(A_{\text{10}}\) represents the Einstein coefficient for the hyperfine 21~cm transition, with a value of \(2.85 \times 10^{-15}\ \text{s}^{-1}\), governing the likelihood of spontaneous emission, 
and \(k_B\) is the Boltzmann constant.
$n_{\rm HI} (x)$ is the neutral hydrogen number density at position $x$ along the line of sight (LoS), $b(x)=(2k_{\rm B}T_{\rm K}(x)/m_{\rm H})^{1/2}$ is the Doppler parameter, in which $m_{\rm H}$ is the hydrogen atom mass, \(T_{\rm K}(x)\) is the gas kinetic temperature, and \(T_{\rm S}(x)\) is the spin temperature, which characterizes the relative population of the two hyperfine energy levels of neutral hydrogen.
$u(\nu)\equiv c(\nu-\nu_{\rm 21~cm})/\nu_{\rm 21~cm}$, where \(\nu_{21\text{cm}} \approx 1420.4\ \text{MHz}\) is the
rest-frame frequency of the 21~cm transition,
and $v(x)$ is the velocity of the hydrogen gas projected to the LoS, including the peculiar velocity and the Hubble flow.

When the optical depth is small, the absorption lines are weak and appear only as subtle fluctuations in the background spectrum. Conversely, a larger optical depth produces deeper and more prominent absorption features, reflecting regions of higher density or lower temperature in the neutral hydrogen gas.
Because the strength and shape of these lines are sensitive to the local density, temperature, and ionization state of the IGM, analyzing the 21 cm forest enables us to infer the small-scale distribution of matter and to trace the thermal evolution of the IGM during the early stages of cosmic history.

The 21~cm forest is complementary to the 21~cm emission line, which is observed as a diffuse signal from neutral hydrogen itself. While the emission signal is typically measured over large areas of the sky using statistical methods such as power spectrum analysis or tomographic imaging, which thereby offers a global view of the IGM during cosmic dawn and the Epoch of Reionization, the 21~cm forest provides a high-resolution, line-of-sight probe of the intervening gas, revealing detailed information about small-scale structures and local physical conditions. Together, these approaches offer a comprehensive picture: the emission line maps the large-scale distribution of neutral hydrogen, whereas the forest targets the finer details of gas density, temperature, and kinematics along a discrete line of sight.

Observations of the 21~cm forest using next-generation radio telescopes, such as the Square Kilometre Array (SKA), are poised to provide transformative insights into cosmic structure formation and reionization\citep[e.g.][]{Ciardi_2015,deLeraAcedo01.2026.SKA}. These studies promise to deepen our understanding of the early universe by probing the distribution and properties of neutral hydrogen along the line of sight\citep[e.g.][]{Bag01.2026.SKA,Barkana01.2026.SKA}.

This chapter provides a comprehensive overview of theoretical modeling, numerical simulations, data-analysis techniques, and observational strategies for studying 21 cm forest and its connection to high-redshift radio sources in the context of future SKA-Low observations.


\begin{figure}
    \centering
    \includegraphics[width=1.0\hsize]{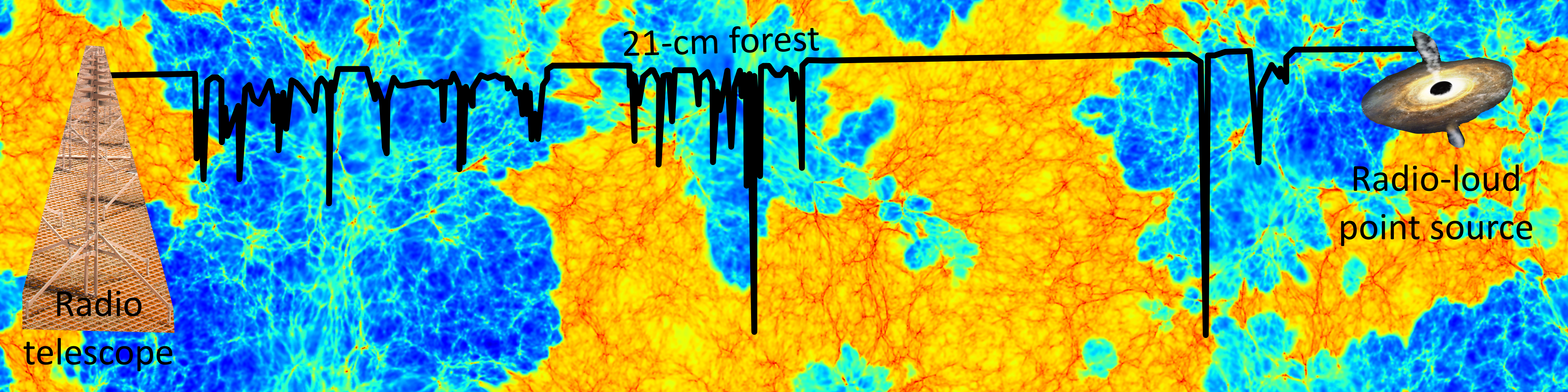}
    \caption{Schematic view of 21~cm forest. Credit: SKAO, Sherwood-relics, International Gemini Observatory, Tomáš Šoltinský.}
    \label{fig:schematic_forest}
\end{figure}

\section{Modelling the 21~cm forest}



\subsection{Analytical vs numerical approach}

Absorption features in the spectra of high-$z$ quasars induced by intervening minihalos and dwarf galaxies have been investigated with a semi-analytic approach by e.g. \cite{Furlanetto_2002}, \cite{Furlanetto_2006}, \cite{Meiksin_2011}, \cite{Xu_2011}, and \cite{Shimabukuro_2014}.
Semi-analytic approaches can provide detailed modeling for small-scale structures that are currently impossible to resolve for numerical hydrodynamical simulations, but can be important for 21~cm forest signals, for which the main contributors of the absorption are small-scale structures.
We need to model the density, temperature, and ionization profiles 
not only for the hydrogen gas within dark matter halos, but also for the infalling gas surrounding halos.
In particular, the cool and dense gas in the infalling region surrounding halos could have a major contribution to the optical depth when the gas has not been sufficiently heated by the X-rays \citep{Xu_2011}. The semi-analytic approach also has the advantage of being able to explore a large parameter space due to its efficient modeling. Similar advantages have fully-analytic models, one of which was developed for the 1D power spectrum of the 21~cm forest by \citet{Shao_2024}.

On the other hand, numerical hydrodynamical simulations combined with radiative (RT) transfer can provide more detailed models of the IGM during the Epoch of Reionization. Examples of numerical simulations which have been used for 21~cm forest studies include \textsc{Aurora} \citep{Pawlik_2017,Bhagwat_2022}, \textsc{CRASH} \citep{Ciardi_2012,Ciardi_2013}, Sherwood-Relics \citep{Puchwein_2023,Soltinsky_2021,Soltinsky_2023} and others \citep{Carilli_2002,Xu_2009, Semelin_2016,Naruse_2024}. From such simulations, we extract, along each line of sight, the hydrogen number density \(n_{\rm H}\), neutral fraction \(x_{\rm HI}\), kinetic and spin temperatures (\(T_{\rm K}\), \(T_{\rm S}\)), and the peculiar velocity of the gas \(v_{\rm pec}\).
With this information we can calculate the 21~cm forest optical depth in discrete form at pixel $i$ in the LOS with $N_{\rm pix}$ pixels in total as
\begin{equation}
\tau_{\mathrm{21~cm}, i} = \frac{3h_{\rm p}c^{3}A_{10} }{32\pi^{3/2}\nu_{\rm 21~cm}^{2}k_{\rm B}} \frac{\delta v}{H(z)} \sum_{j=1}^{N_{\rm pix}}\frac{n_{{\rm HI}, j}}{b_{j}T_{{\rm S}, j}}\exp\left( - \frac{ (v_{{\rm H},i}-u_{j})^{2}} {b^{2}_{j}}\right), \label{eq:tau21_discrete}
\end{equation}
\noindent
where $\delta v$ is the velocity width of the pixels, $H(z)$ is the Hubble parameter, $n_{\rm HI}=n_{\rm H}x_{\rm HI}$, $u_{j}=v_{{\rm H},j}+v_{{\rm pec},j}$ and $v_{\rm H}$ is the Hubble velocity. However, most simulations include only the ionizing (UV) radiation that drives reionization, 
while neglecting the contribution of X-ray photons. 
The latter, however, play a crucial role in determining the thermal state of the IGM and hence the 21~cm forest signal. 
Owing to their long mean free path,

\begin{equation}\label{eq:Xray_mfp}
\lambda_{\rm X} \simeq 1~{\rm cMpc}\,
x_{\rm HI}^{-1}(1+\delta)^{-1}
\left(\frac{E_{\mathrm{X}}}{0.2\,{\rm keV}}\right)^3
\left[\frac{1+z}{10}\right]^{-2},
\end{equation}
where $\delta$ is the matter overdensity contrast and $E_{\rm X}$ is the energy of the X-ray photon. 
Since the mean free path increases rapidly with photon energy ($\lambda_{\rm X} \propto E_{\rm X}^3$) 
and decreases in dense or neutral regions, X-ray photons can penetrate far beyond local ionized bubbles, 
heating large volumes of the IGM. 
Therefore, their effect is often approximated as a spatially homogeneous background, 
whose luminosity can be parameterized as \citep{Furlanetto_2006b}
\begin{equation}\label{eq:Xray_luminosity} 
L_{\rm X}=3.4\times 10^{40}\rm\,erg\,s^{-1} \, \mathnormal{f_{\rm X}} \left(\frac{SFR}{1\,M_{\odot}\rm\,yr^{-1}}\right),
\end{equation}
\noindent
where the $f_{\rm X}$ is the X-ray background radiation efficiency and the SFR is the star formation rate, which preheats the IGM in the post-processing of cosmological simulations, rather than following the more computationally expensive RT of the photons. Differently from analytic approaches, the numerical simulations mentioned above are computationally expensive, and hence it is difficult to explore a large parameter space with them.

An alternative approach that requires less computational resources is to utilize semi-numerical tools. For example, the \textsc{21cmFAST} code (\citealt{Mesinger_2011,Murray_2020}, but see also \citealt{Mack_2012}) has been widely used to vary the X-ray background radiation, as well as the reionization and dark matter models and to investigate the impact on the 21~cm forest \citep[e.g.][]{Ewall_Wice_2014,Thyagarajan_2020,Shao_2023,Soltinsky_2025}.
Combining the semi-numerical tools with the semi-analytic modeling, a high enough dynamic range can thus be reached to model in detail both the large-scale environment and the small-scale physics driving the signal \citep{Shao_2023}. 

\subsection{Effect of assumptions implemented in the modelling}\label{sec:assumptions_effect}

\begin{figure}
    \begin{minipage}{\linewidth}
 	  \centering
 	  \includegraphics[width=\linewidth]{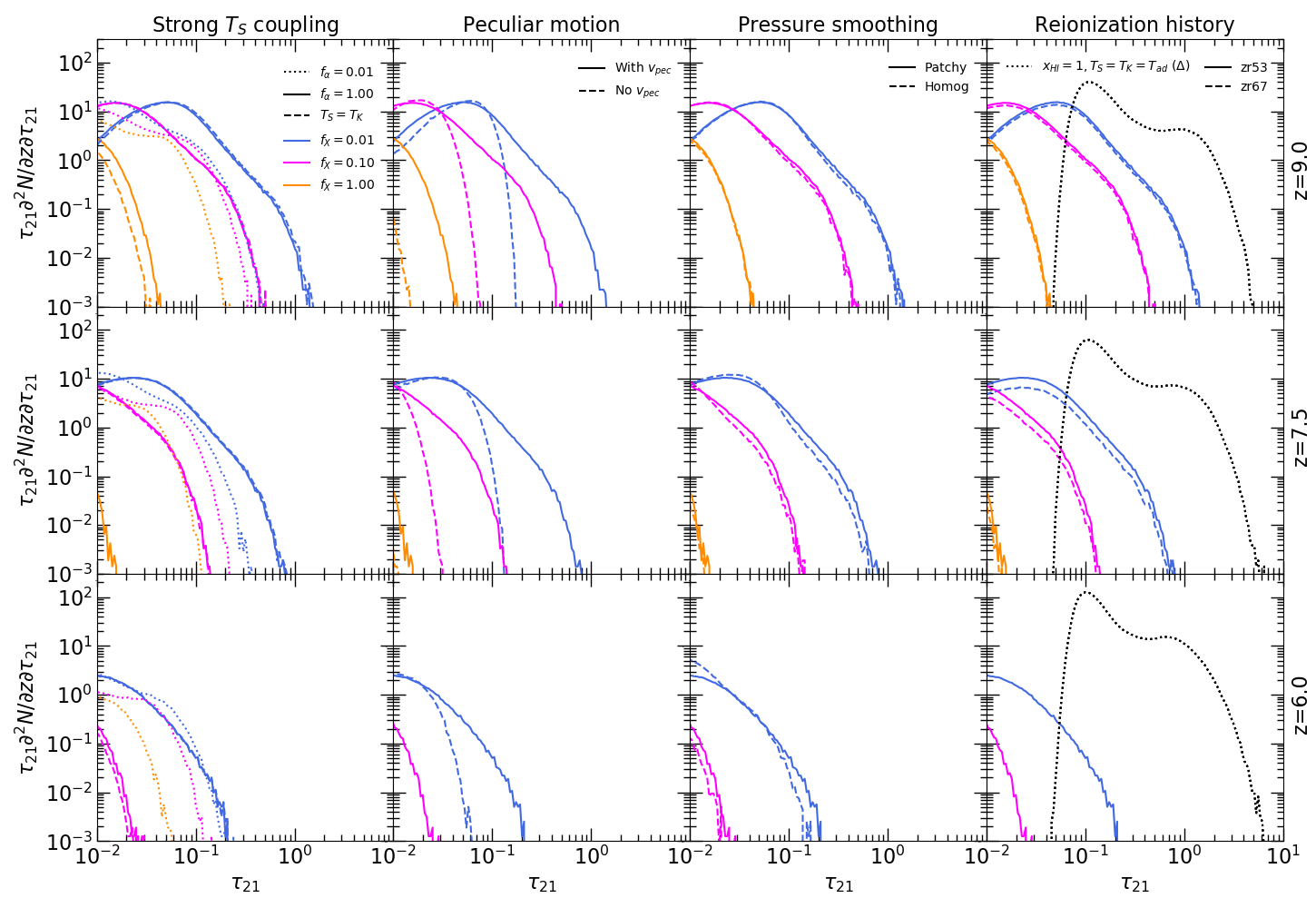}
	\end{minipage}
	\vspace{-0.3cm}
    \caption{The differential number density of synthetic 21~cm forest absorption features at $z=9$ (top row panels), $7.5$ (middle row panels) and $6$ (bottom row panels). Each panel shows cases in which the X-ray efficiencies, $f_{\rm X}=0.01$ (blue curves), $f_{\rm X}=0.1$ (fuchsia curves) and $f_{\rm X}=1$ (orange curves) is assumed. From left to right various effects on the 21~cm forest are tested, namely strength of Ly$\alpha$ coupling of the $T_{\rm S}$, redshift space distortions, topology and timing of reionization. This figure is taken from \citet{Soltinsky_2021}.} 
    \label{fig:AoFdiff}
\end{figure}

In the previous section, we discussed various analytical, semi-analytical, and numerical approaches that have been used to model the 21 cm forest, each with its own advantages and limitations. Despite their methodological differences, these models are generally built upon a set of simplifying assumptions regarding the thermal and ionization state of the intergalactic medium and the relevant radiative processes. In what follows, we describe some of these assumptions and how they affect the detectability of the 21~cm forest signal. 

During the end stages of the reionization the universe is expected to be filled with a homogeneous Ly$\alpha$ photons background strong enough for the spin temperature to be coupled to the gas kinetic temperature via the Wouthuysen-Field effect \citep{Wouthuysen_1952,Field_1958}. Hence, typically it is assumed that $T_{\rm S}=T_{\rm K}$, which allows us to avoid the complex calculation of $T_{\rm S}$. However, this is not necessarily the case, and in fact the spin temperature could also be partially coupled to the radio background temperature. As a result, the 21~cm forest absorption can be modestly increased or decreased in comparison to a $T_{\rm S}=T_{\rm K}$ case, depending on the redshift evolution of $T_{\rm K}$ \citep{Semelin_2016,Soltinsky_2021}. This is shown in the leftmost column panels in Fig.~\ref{fig:AoFdiff}, where the distribution of the absorption lines in the simulated 21~cm forest spectra is shown when the $T_{\rm S}=T_{\rm K}$ approximation was assumed (dashed curves), a full computation of $T_{\rm S}$ was performed (solid curves) and weak Ly$\alpha$ coupling was considered (dotted curves).

Another common approximation is to omit the effect of redshift space distortions (RSD), which can increase the optical depth of individual absorption features and clusters them together. As a consequence, the inclusion of RSD can boost the high $\tau_{21}$ tail of the distribution of absorption features by up to an order of magnitude \citep{Semelin_2016,Soltinsky_2021} as we can see in the middle left column of Fig.~\ref{fig:AoFdiff}, as well as the 1D power spectrum at small scales \citep{Soltinsky_2025}. We note that hydrodynamical simulations naturally produce the gas peculiar velocity fields, $v_{\rm pec}$, which can then be used to incorporate the effect of RSD into the 21~cm forest modeling. 

On the other hand, \citet{Soltinsky_2021} found that the hydrodynamical response of the gas to the inhomogeneous (patchy) reionization has a modest effect (see the middle right column panels in Fig.~\ref{fig:AoFdiff}). The structures in the neutral islands of the IGM which have not yet experienced reionization (and thus Jeans pressure smoothing) produce 21~cm forest absorption features which are only slightly stronger than those in a case of homogeneous reionization.

Besides the morphology of the reionization, the timing can affect the detectability of the 21~cm forest too. For example, in the rightmost middle panel of Fig.~\ref{fig:AoFdiff}, which corresponds to $z=7.5$, the model in which the reionization is completed at $z_{\rm r}=6.7$ (dashed curves) has slightly suppressed signal relatively to the model with $z_{\rm r}=5.3$ (solid curves) because the reionization has progressed faster and more IGM is ionized and hot. For comparison a case of no reionization and no X-ray heating (i.e. $T_{\rm S}=T_{\rm K}=T_{\rm ad}=2.73\rm\, K\ (1+\delta)^{2/3}(1+z)^2/(1+z_{\rm dec})$, where $T_{\rm ad}$ is the adiabatic temperature, $z_{\rm dec}=147.8$ \citep{Furlanetto_Oh_2006}, and $x_{\rm HI}=1$) is indicated by dotted black curves. Recent observations of the large spatial fluctuations in the Ly$\alpha$ forest opacity \citep{Becker_2015,Bosman_2022}, the presence of damping wings in the Ly$\alpha$ forest spectra from neutral islands at $z<6$ \citep{Spina_2024,Becker_2024,Zhu_2024,Sawyer_2025}, Ly$\alpha$ equivalent widths \citep{Nakane_2024}, thermal widths of Ly$\alpha$ forest transmission spikes at $z>5$ \citep{Gaikwad_2020}, clustering of Ly$\alpha$ emitters \citep{Weinberger_2019}, deficit of Ly$\alpha$ emitting galaxies around extended Ly$\alpha$ absorption troughs \citep{Kashino_2020,Keating_2020,Christenson_2021}, long dark gaps in the Ly$\alpha$ forest \citep{Zhu_2021} and Ly$\beta$ forest \citep{Zhu_2022}, mean free path of ionizing photons at $z=6$ \citep{Becker_2021,Cain_2021,Zhu_2023,Gaikwad_2023}, the cross-correlation of the Ly$\alpha$ forest-[\hbox{O$\,\rm \scriptstyle III$}] emitters \citep{Kakiichi_2025} and the cosmic microwave background measurement of the electron scattering optical depth  \citep{Pagano_2020}
are consistent with the reionization not being completed before $z\sim6$ \citep[e.g.][]{Kulkarni_2019,Nasir_2020,Choudhury_2021,Qin_2021}. For alternative explanations and the evidence against this see \citet{DAloisio_2015}, \citet{Davies_2016}, \citet{Chardin_2017}, \citet{Meiksin_2020} and \citet{Zheng_2026}. However, if the late-end reionization scenario is indeed correct, we can expect large islands of neutral hydrogen to persist until $z\sim6$ \citep[e.g.][]{Lidz_2007,Mesinger_2010}, and hence the 21~cm forest absorption might be detectable at such low redshift if the gas is not preheated above $T_{\rm S}\sim100\,\rm K$ \citep{Soltinsky_2021}.

Furthermore, one of the open questions in the field is how much do minihalos contribute to the 21~cm forest absorption \citep{Furlanetto_2006,Xu_2011,Kadota_2023,Naruse_2024}, as minihalos are difficult to model in numerical simulations due to the large dynamic range required. Indeed, large simulation boxes are needed to capture the rarest and most massive halos, but at the same time the modeling of minihalos requires fine resolution. Therefore, most studies employing cosmological simulations focus on the 21~cm forest originating from the diffuse IGM only \citep[e.g.][]{Ciardi_2013,Soltinsky_2021}. Additionally, feedback processes may easily suppress the absorption signature from minihalos \citep{Meiksin_2011,Park_2016,Nakatani_2020}. It has been shown, though, that the presence of minihalos impacts the evolution of reionization itself \citep{Ciardi_2006,Yue_2009,Chan_2024}, and therefore can indirectly affect also the 21~cm forest.

Finally, often the effect of the background radio-loud quasar radiation on the proximate 21~cm forest is omitted. By postprocessing cosmological hydrodynamical/RT simulations with a multifrequency 1D RT simulations incorporating quasar spectrum from the UV to X-ray photons to model the impact of the quasar radiation on its surroundings, \citet{Soltinsky_2023} found that not only the signal is completely suppressed close to the quasar due to the \HII~bubble pre-ionized by the UV radiation from the galaxies populating this region, but can be partially suppressed far away from the quasar because of the heating driven by the X-ray photons emitted by the quasar itself. It should be noted that, because of their long mean free path, X-ray photons penetrate the IGM beyond the ionization front far away from the background radio-loud quasar.

\section{Evaluating 21~cm forest with the SKA}
\subsection{Direct observation of spectra \& sensitivity}


Previous works propose a direct counting method to detect the 21~cm absorption lines imprinted by cold neutral hydrogen in and around minihalos and dwarf galaxies along the sightlines to bright, high-redshift radio sources \citep[e.g.][]{Furlanetto_2002,Furlanetto_2006,Xu_2011}. In this approach, the optical depth of a minihalo is determined by the neutral hydrogen number density, the spin temperature of the gas, and the line profile function that describes the velocity dispersion of the gas. The optical depth depends on the impact parameter, which corresponds to the maximum distance along the line of sight at which significant absorption occurs.

By modeling the halo mass function and the internal and ambient gas properties of each minihalo, the cumulative number of absorption features per unit redshift is given by

\[
\frac{dN(>\tau)}{dz} = \frac{dr}{dz} \int_{M_{\mathrm{min}}}^{M_{\mathrm{max}}} \frac{dN}{dM}\,\pi\,r_\tau^2(M,\tau)\,dM,
\]
where \(\frac{dr}{dz}\) is the comoving line element, \(\frac{dN}{dM}\) is the halo mass function, and \(r_\tau(M,\tau)\) is the maximum impact parameter for which the optical depth exceeds \(\tau\).

This direct counting strategy requires the high sensitivity and kHz-level frequency resolution of instruments to detect weak absorption signals (with \(\tau \sim 0.01\)–0.1). Ultimately, the statistical distribution of these 21~cm absorption lines provides a powerful tool to probe small-scale cosmological fluctuations and to constrain key parameters such as the neutrino mass, the running spectral index, and dark matter particle mass\citep{Shimabukuro_2014}. Similarly, \citet{Soltinsky_2021} have used the direct counting strategy for the 21~cm absorption lines originating from the diffuse IGM rather than minihalos. They also suggest that a null-detection of these features in observations by SKA1-low or SKA2-low can be translated to a lower limit on the X-ray background radiation.

\subsection{Statistical measurements \& sensitivity}

In statistical analysis, the brightness temperature of the 21 cm forest. 
The brightness temperature includes not only information about the optical depth but also contains information about the background radio source. 
This can be expressed as
\begin{equation}
\delta T_b(\hat{s}, \nu) \approx \frac{T_S(\hat{s}, z) - T_\gamma(\hat{s}, z)}{1 + z} \, \tau(\hat{s}, z),
\label{Eq_21cm_diff_brightness_temp}
\end{equation}
where $\hat{s}$ denotes the unit vector along the line of sight (i.e., the observing direction), 
$T_S$ is the spin temperature of neutral hydrogen, and 
$T_\gamma$ is the brightness temperature of the background radiation, which is related to the flux density of the background radio source.

Compared with direct detection, statistical analysis methods, such as increased variance \citep{Mack_2012}, power spectrum \citep{Thyagarajan_2020,Shao_2023}, wavelet transform \citep{Shimabukuro_2025a} and topological data analysis\citep{2026PhRvD.113h3525S}, have significantly improved the observability of the 21 cm forest. First of all, statistical methods effectively suppress noise through signal superposition. Even if individual absorption lines are difficult to distinguish, they can still extract the overall statistical characteristics and significantly improve the signal-to-noise ratio. Secondly, they are less sensitive to instrumental calibration errors. By analyzing the spatial or frequency correlation of signals, the impact of systematic errors can be reduced. In addition, statistical methods have relatively relaxed requirements for data resolution. Even if the observational resolution is limited, they can still constrain the signal characteristics by measuring the amplitude of fluctuations on different scales.

\subsubsection{Power spectrum}\label{sec:power_spectrum}

The power spectrum is one of the most commonly used statistical analysis methods in cosmology and astronomy\cite[e.g.][]{Acharya01.2026.SKA}. It quantitatively describes the fluctuation characteristics of signals at different scales through Fourier transform. In the study of the 21 cm forest, \citet{Thyagarajan_2020} first proposed the application of one-dimensional (1D) power spectrum along the line of sight, and it has been shown that this 1D power spectrum analysis can effectively suppress the impact of cosmic variance and significantly improve the signal-to-noise ratio of the observed signal. \citet{Shao_2023} further found that the 1D power spectrum of the 21 cm forest has an important scientific value, capable of breaking the degeneracy between dark matter properties and intergalactic medium temperature in traditional observations, thereby achieving simultaneous constraints on the mass of dark matter particles and the history of cosmic heating. Similarly, when the 1D power spectrum is computed from the transmitted 21~cm forest flux, \citet{Soltinsky_2025} have found that the measurement of this statistical observable can be used to constrain both the thermal and ionization state of the high-$z$ IGM even with an observation of a few radio-background sources. These groundbreaking discoveries provide a new theoretical framework for studying early universe physics using the 21 cm forest.

The observed frequency-dependent brightness temperature $\delta T_b(\hat{s}, \nu)$ can also be represented in terms of the line-of-sight distance $r_z$, as $\delta T_b(\hat{s}, r_z)$. The Fourier transform of $\tilde{\delta T'_b}(\hat{s}, r_z)$ is defined as:
\begin{equation}
\delta \tilde{T}' \left( \hat{s}, k_{\parallel} \right) = \int \delta T_b' \left( \hat{s}, r_z \right) e^{-i k_{\parallel} r_z} \, dr_z.
\end{equation}
The corresponding one-dimensional power spectrum is
\begin{equation}
P \left( \hat{s}, k_{\parallel} \right) = \left| \delta \tilde{T}' \left( \hat{s}, k_{\parallel} \right) \right|^2 \left( \frac{1}{\Delta r_z} \right).
\end{equation}
The factor $1/\Delta r_z $ serves as the normalization factor, with $\Delta r_z$ denoting the length of the sightline in question. By extracting multiple segments of neutral regions from the spectra of different radio sources, we can obtain the expected value of the one-dimensional power spectrum $ P\left(k_{\parallel}\right) \equiv \langle P\left(\hat{s}, k_{\parallel}\right) \rangle$.

The uncertainties in 21 cm forest observations primarily stem from the thermal noise of the interferometer array and the sample variance in power spectrum measurements. The sample variance of the one-dimensional power spectrum can be expressed as $P^S = \sigma_P(k)/\sqrt{N_s \times N_m}$, where $\sigma_P(k)$ is the standard deviation of $P(k)$, $N_s$ represents the number of independent measurements on different neutral patches of $\Delta r_z$, and $N_m$ is the number of independent modes in each $k$-bin from each measurement. Regarding the thermal noise error, \citet{Shao_2023} follows the approach taken by \cite{Thyagarajan_2020}, assuming each spectrum is measured twice separately, or the total integration time is divided into two halves, and the cross-power spectrum is practically measured to avoid noise bias. Under this measurement scheme, the integration time for each measurement of the spectrum is $\delta t_{0.5} = 0.5 \delta t$, resulting in an increase in the thermal noise on the spectrum by a factor of $\sqrt{2}$. Then, the uncertainty in the 1D power spectrum due to thermal noise is given by

\begin{equation}
P^N = \left( \frac{\lambda_z^2 T_{\text{sys}}^2}{A_{\text{eff}}} \right) \left( \frac{\Delta r_z}{2 \Omega \Delta \nu_z \delta t_{0.5}} \right) 
\end{equation}
where $A_{\text{eff}}/T_{\text{sys}}$ represents the sensitivity of the telescope array, $\Omega = \pi(\theta/2)^2$ is the solid angle of the telescope beam, $\theta$ denotes the angular resolution, and $\Delta \nu_z$ is the total observing bandwidth corresponding to $\Delta r_z$. For $N_s$ measurements, the thermal noise $P^N$ will further decrease by a factor of $\sqrt{N_s}$.

\begin{figure}
    \centering
    \includegraphics[width=0.8\hsize]
{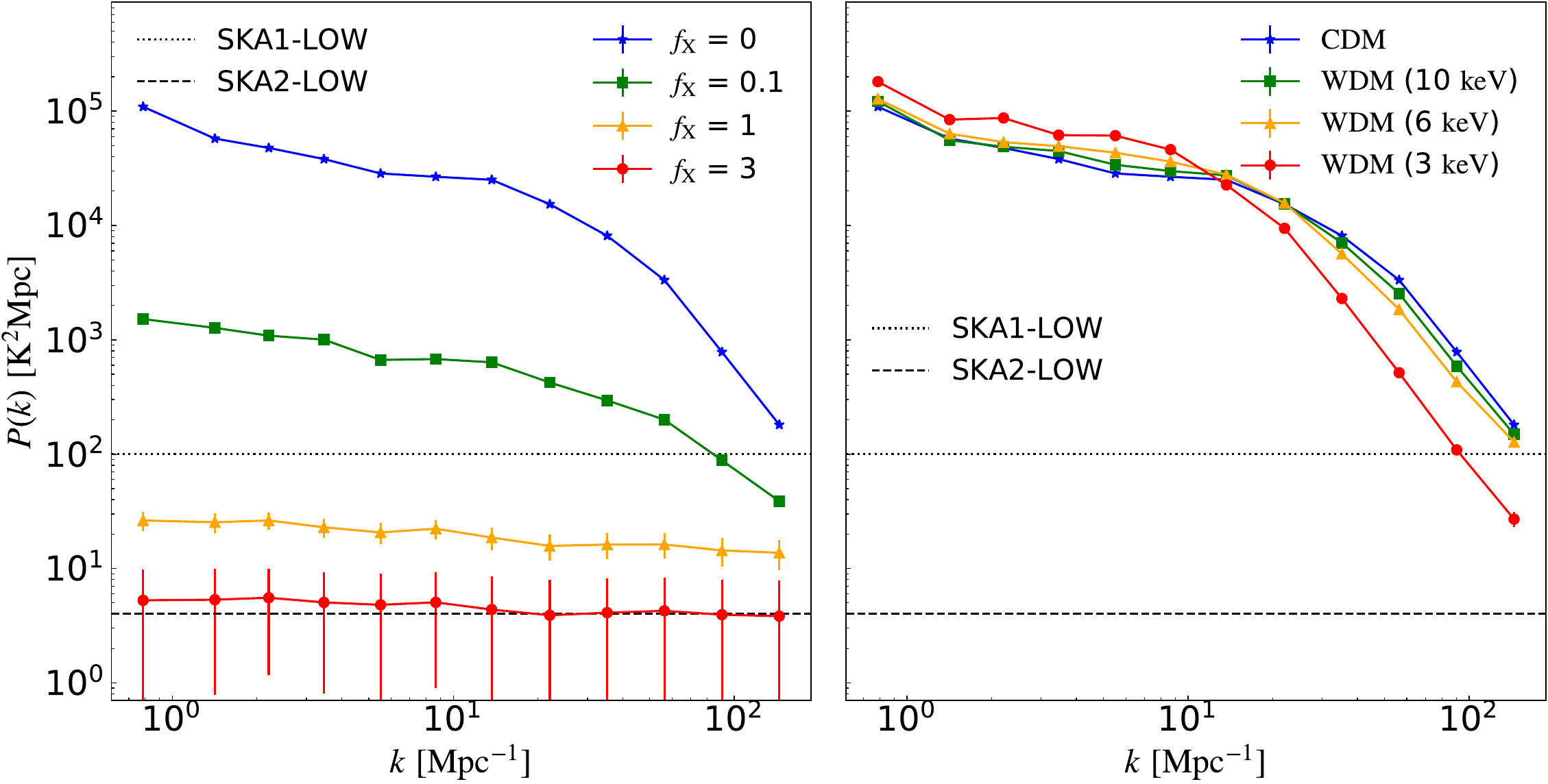}
    \caption{The expected 1D power spectrum of the 21 cm forest for different $f_{\rm X}$ (left panel) and $m_{\mathrm{WDM}}$ (right panel), with $S_{150}$ = 10 mJy. This figure is taken from~\citet{Shao_2023}.}
  \label{fig:Pk}
\end{figure}

Figure~\ref{fig:Pk} shows the 1D power spectra that we have simulated. The left panel displays the 1D power spectra corresponding to different heating histories in the CDM model, where $f_{\rm X}$ = 0, 0.1, 1, and 3 represent different heating histories. As \(f_{\rm X}\) increases, the temperature of the IGM rises, leading to a significant suppression of the power spectrum across all scales. The right panel illustrates the 1D power spectra for the CDM model without heating and for WDM models with particle masses of 10 keV, 6 keV, and 3 keV, respectively. A lower \(m_{\rm WDM}\) reduces small-scale density fluctuations, thereby suppressing the small-scale power spectrum. Unlike heating effects, the influence of dark matter properties is primarily reflected in changes to the slope of the power spectrum, rather than a significant suppression of the overall amplitude. The dotted lines and dashed lines represent the thermal noise corresponding to the sensitivities of the SKA1-LOW and SKA2-LOW, respectively, for which we have adopted parameters of \( A_{\text{eff}}/T_{\text{sys}} = 800 \, \text{m}^2/\text{K} \) and \( 4000 \, \text{m}^2/\text{K} \). Additionally, \citet{Shao_2023} has assumed an integration time of 100 hours on each source, a spectral resolution of 1 kHz, and 100 observed neutral segments. Because heating and dark matter effects leave distinct imprints on the 21 cm forest 1D power spectrum, the two can be disentangled, allowing us to simultaneously probe the thermal history of the Universe and the properties of dark matter.

\subsubsection{Wavelet Scattering Transform}
\citet{Shimabukuro_2025a} introduced the wavelet scattering transform (WST) to study the 21~cm forest beyond the power spectrum. WST is a multi-scale method that decomposes a signal into scale-dependent coefficients and is well suited to complex data with nested structures and non-Gaussian features. Applied to the 21~cm forest brightness temperature, it captures localized fluctuations and cross-scale interactions through wavelet filters of different sizes. Its nonlinear modulus and averaging operations also make it stable against small perturbations and robust to instrumental and astrophysical noise. The resulting coefficients include first-order terms (\(S_1\)), which trace intensity variations at individual scales, and second-order terms (\(S_2\)), which quantify hierarchical coupling between scales.

The WST first convolves the 21~cm forest brightness temperature spectrum $\delta T_b$ with a family of wavelets $\psi_{j,l}(x)$. In the WST, the scale index $j$ determines the dilation of the wavelet filter, that is, how much the wavelet is stretched or compressed in scale, while the orientation index $l$ specifies the direction of the wavelet. Small $j$ corresponds to fine structures or rapid oscillations (small scales), whereas large $j$ captures broad, slowly varying patterns (large scales). Different $l$ values allow the WST to extract directional features from various orientations in the data. However, $l$ is redundant in one-dimensional 21~cm forest spectrum data.

The hierarchical coefficients start with the spatial mean
\begin{equation}
S_0 \;=\; \langle \delta T_b \rangle ,
\end{equation}
which encodes global information such as the average brightness temperature.  Localised fluctuations at scale $j_1$ are captured by the first-order coefficients obtained after convolution, modulus, and spatial averaging,
\begin{equation}
S_1^{j_1} \;=\; \Bigl\langle \, \bigl| \, \delta T_b * \psi_{j_1}(x) \bigr| \Bigr\rangle .
\end{equation}
To probe cross-scale structure, the modulus field $\bigl| \delta T_b * \psi_{j_1} \bigr|$ is convolved with a second wavelet of scale $j_2$; taking the modulus again and averaging yields the second-order coefficients
\begin{equation}
S_2^{j_1,j_2} \;=\; \Bigl\langle \, \Bigl| \, \bigl| \delta T_b * \psi_{j_1}(x) \bigr| * \psi_{j_2}(x) \Bigr| \Bigr\rangle ,
\end{equation}
which quantify how structures at scale $j_1$ modulate those at scale $j_2$ and thus encode higher-order, non-Gaussian statistics.

\begin{figure}
    \centering
    \includegraphics[width=1.0\hsize]{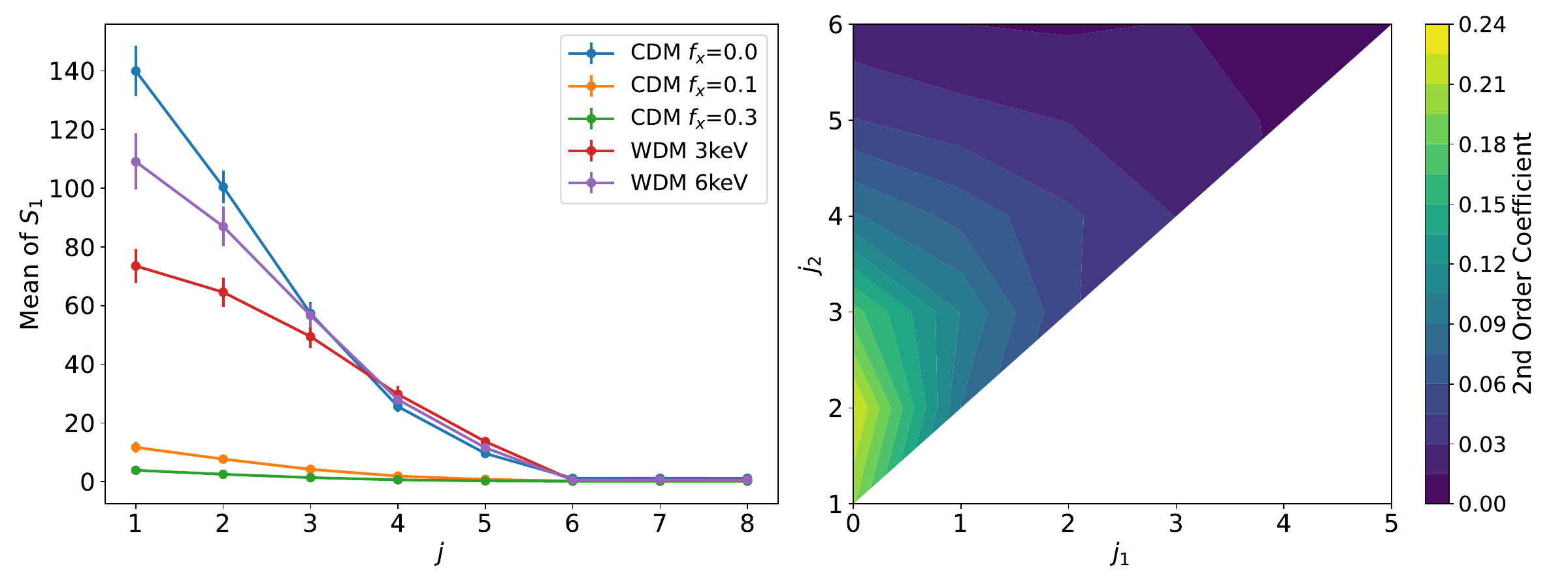}
    \caption{({\it left}) first-order WST coefficient for CDM model, CDM + $f_{\rm X}=0.1,0.3$ models and Warm Dark Matter(WDM) models with different WDM mass. ({\it right}) second-order WST coefficient for CDM model. This figure is taken from \citet{Shimabukuro_2025a}.}
    \label{fig:WST}
\end{figure}

In Fig.\ref{fig:WST}, we show the first and second orders of the WST coefficient. The left panel shows the scale‐dependent first‐order scattering coefficients, $\langle S_1 \rangle$, as a function of the wavelet scale $j(=j_1)$ for five cosmological models: cold dark matter (CDM) with $f_{\rm X} = 0.0$, $0.1$, and $0.3$, and warm dark matter (WDM) with particle masses of 3\,keV and 6\,keV. In all cases, $\langle S_1 \rangle$ falls off rapidly with increasing $j$, indicating that small‐scale fluctuations dominate the WST response; models with more free‐streaming (higher $f_{\rm X}$ or lighter WDM) yield systematically lower amplitudes at each $j$. The right panel presents the second‐order scattering coefficients, $S_2$, as a function of the scale pair $(j_1, j_2)$. The strongest cross‐scale coupling occurs at the smallest scales (low $j_1$ and $j_2$), and $S_2$ declines steeply as either index increases, reflecting the decreasing interaction between widely separated scales.

\subsubsection{Sample size and statistical problem}

\begin{figure}[t]
    \centering
    \includegraphics[width=\textwidth]{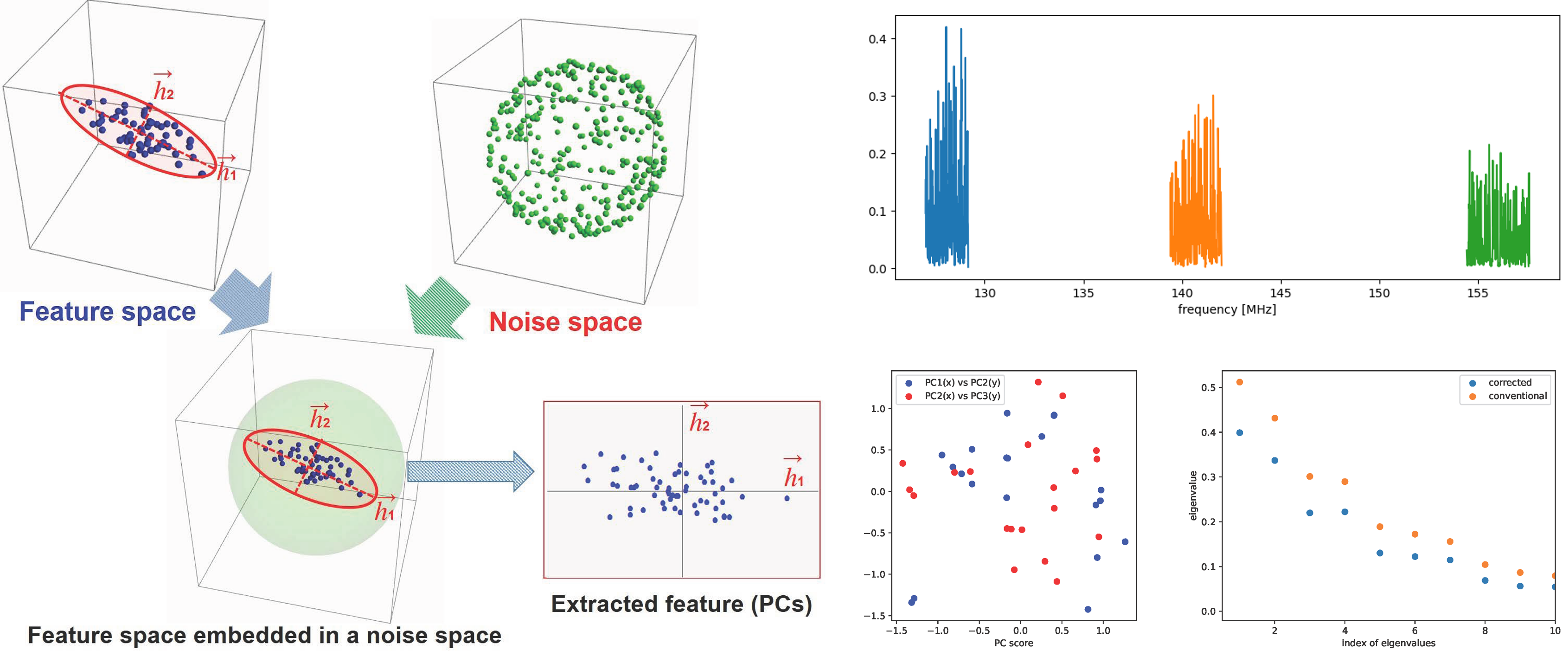}
    \caption{A schematic description of the high-dimensional principal component analysis (PCA) of the 21~cm forest. 
    Left panel: 
    Structure of the HDLSS data and their analysis.
    Even if we have significant important features, because of the existence of a huge noise sphere, they are completely embedded in the observed data. 
    The high-dimensional PCA can subtract the noise sphere and enables to extract the desired features 
    \citep{Takeuchi_2024}.  
    Right panel: 
    Application to the simulated 21~cm forest. 
    Top panel shows the optical depth of the realization of the 21~cm forest based on the Illustis TNG simulation.
    Bottom-left panel shows the result of the high-dimensional PCA applied to the simulated 21~cm forest data.
    Bottom-right panes compares the eigenvalues (contributions) obtained from the classical and high-dimensional PCA.}
    \label{fig:HI_forest}
\end{figure}

As discussed above, the 21~cm forest is expected to provide a powerful probe of large-scale structure and galaxy formation in the high-$z$ Universe.
However, even with the SKA, the number of sufficiently bright background radio sources at such redshifts is expected to be limited \citep[][and references therein]{Takeuchi_2016}.
The total number of usable sightlines may reach only a few tens to $\sim 100$ over the full survey.
In contrast, the high frequency resolution of the SKA yields a very large spectral data dimension.
Thus, the observational dataset naturally falls into a regime where the sample size $n$ is much smaller than the data dimension $d$.

Consider a $d$-dimensional parent population with covariance matrix $\tilde{\Sigma}_d > \tilde{O}$ and eigenvalues $\lambda_1 \ge \lambda_2 \ge \dots \ge \lambda_d > 0$. 
From this population we draw $n$ independent samples $\vec{x}_1,\dots,\vec{x}_n$ ($d>n$) and construct the data matrix $\tilde{X} = (\vec{x}_1,\dots,\vec{x}_n)$.
The sample covariance matrix is $\tilde{S} = \dfrac{1}{n}\tilde{X}\tilde{X}^\top$, which becomes rank-deficient when $n \ll d$.
In practice, principal component analysis (PCA) is often used to extract dominant structures, but in this regime the leading eigenvalues can be strongly biased by high-dimensional noise.

A useful correction is obtained by considering the dual representation $\tilde{S}_{\rm D} = \frac{1}{n}\tilde{X}^\top \tilde{X}$ and modifying the sample eigenvalues $\hat{\lambda}_i$ as
\begin{align}
    \check{\lambda}_i
    =
    \hat{\lambda}_i
    -
    \frac{1}{n-i}
    \left(
        \mathrm{tr}\,\tilde{S}_{\rm D}
        -
        \sum_{j=1}^i \hat{\lambda}_j
    \right)
    \quad (i=1,\dots,n-1) .
    \label{eq:eigenvalues_estimation_short}
\end{align}
This estimator removes the high-dimensional noise contribution and provides a more stable characterization of physically meaningful variance components.

To illustrate this point, we constructed a mock sample of 21~cm forest spectra from the Illustris TNG simulation\footnote{\url{https://www.tng-project.org/media/}.}.
Even when significant physical features are present, they can be masked by high-dimensional noise in conventional PCA.
Figure~\ref{fig:HI_forest} shows that the corrected analysis enables more robust extraction of such features, which will be important for exploring the spatial distribution and evolution of the 21~cm forest in the SKA era.

\subsubsection{Practical limitations: frequency-dependent systematics and validation}
The sensitivity estimates discussed above can be optimistic if they neglect frequency-dependent instrumental and analysis systematics (e.g., residual bandpass structure, beam chromaticity and the associated sidelobe-induced mode-mixing of foreground emission, and RFI-related spectral structure\citep[e.g.][]{Burba01.2026.SKA}). Such effects are known to complicate low-frequency 21~cm analyses with precursor/pathfinder facilities (e.g., GMRT/uGMRT, MWA, LOFAR, and HERA; \citealt{Paciga_2013,Beardsley_2016,EwallWice_2016,Patil_2017,Mertens_2020,Abdurashidova_2022,Abdurashidova_2023})
. Mitigating these effects will require improved bandpass/beam characterization and end-to-end validation tests (e.g., injection and null tests). While the brightest targets may, in principle, enable high-spectral-resolution follow-up, we emphasize that in the near term (early commissioning and science verification) the most robust SKA results are expected to come from large-scale spectral statistics of the 21~cm forest. Such measurements can constrain the mean 21~cm optical depth and hence place lower limits on the IGM spin temperature, as well as provide coarse constraints on the reionization history \citep[e.g.,][]{Thyagarajan_2020}.

\section{Physics from the 21~cm forest}

\subsection{Astrophysics}

\begin{figure}
    \begin{minipage}{\linewidth}
 	  \centering
 	  \includegraphics[width=0.5\linewidth]{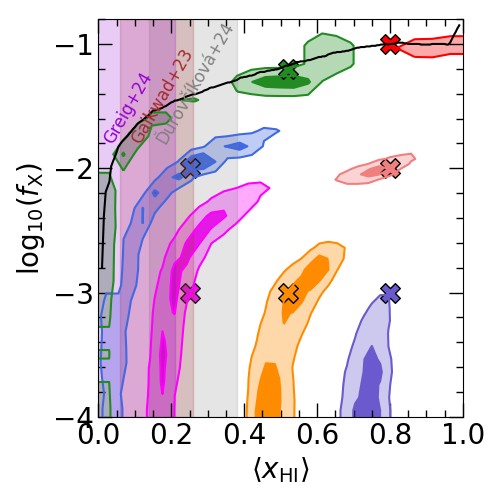}
	\end{minipage}
    \vspace{-0.3cm}
    \caption{The posterior distributions of $\mathrm{log}_{10}f_{\rm X}$ and $\langle x_{\rm HI}\rangle$ from a Bayesian statistics analyses based on 1D power spectrum of 21~cm forest corresponding to 1 and 2$\sigma$ confidence regions. True values of the $\mathrm{log}_{10}f_{\rm X}$ and $\langle x_{\rm HI}\rangle$ are indicated by the crosses with corresponding colour to the contours. This figure was taken from \citet{Soltinsky_2025} and modified.}
    \label{fig:1DPS_MCMC}
\end{figure}

The hot and ionized bubbles of the IGM have been explored extensively through the observations of the Ly$\alpha$ forest signal. On the other hand, the studies of the cold and neutral islands of the intergalactic gas lag behind. Fortunately, the 21~cm forest signal has a potential to be one of the few probes of such regions in the IGM. Given its sensitivity to the kinetic and spin temperature of the IGM (see. Eq.~\ref{eq:tau21} and~\ref{eq:tau21_discrete}) it has been suggested that this signal can be used to constrain the thermal state of the neutral IGM at $z\geq6$ and the radiation that pre-heats these regions including the X-ray background radiation \citep{Xu_2009,Xu_2011,Ewall_Wice_2014,Shao_2023}. An example of such constraining power, which was already mentioned in Sec.~\ref{sec:power_spectrum}, is presented in Fig.~\ref{fig:1DPS_MCMC} for which mock observations of 1D power spectrum of the 21~cm forest from 10 PSO J0309+27-like quasars \citep{Belladitta_2020} each over $50\,\rm hr$ by the SKA1-low was assumed. However, if the neutral gas is pre-heated to $\gtrsim100\,\rm K$, the 21~cm forest absorption is suppressed to $<1\%$ leaving the signal likely undetectable \citep{Ciardi_2013,Soltinsky_2021}. Given this, a null-detection of the 21~cm forest signatures can be translated to lower limits on the X-ray driven pre-heating \citep{Mack_2012,Soltinsky_2021,Soltinsky_2025}. For instance, if the observations described above were to result in a null detection, lower limits on $f_{\rm X}$ defined by the solid black curve in Fig.~\ref{fig:1DPS_MCMC} can be acquired. Such results would be important as the X-ray background radiation and the history of its sources are still largely unconstrained.

Recent JWST observations provide relevant context for this uncertainty.
Deep JWST spectroscopy has revealed a higher-than-expected incidence of AGN signatures among galaxies at $z\gtrsim 4$--7, including faint broad-line (type-1) AGNs identified with NIRSpec \citep[e.g.,][]{Harikane_2023},
and has also reported compelling evidence for accreting black holes in very high-redshift galaxies \citep[e.g.,][]{Maiolino_2024,Goulding_2023}.
At the same time, many JWST-identified AGN candidates are generally undetected or only weakly constrained in X-rays, suggesting that the emerging high-$z$ AGN population does not yet translate into a well-determined early X-ray background \citep[e.g.,][]{Maiolino_2025,Mazzolari_2024}.
This motivates treating the amplitude and spectral properties of X-ray preheating (often summarized by $f_{\rm X}$) as an uncertain but testable ingredient, and highlights the 21~cm forest as an independent probe of the neutral-IGM thermal history.

In addition, the ionization state of the IGM, and hence the reionization history, can be probed with the 21~cm forest too \citep{Xu_2009,Soltinsky_2025}. Such constraints would be complementary to numerous measurements based on the Ly$\alpha$ forest \citep[e.g.][]{Gaikwad_2023,Durovcikova_2024,Greig_2024} and CMB \citep[e.g.][]{Pagano_2020} observations as we can see in Fig.~\ref{fig:1DPS_MCMC}.

The 21~cm forest absorption features can be suppressed by not only the background X-ray radiation but also by the X-ray photons emitted by the background radio-loud quasar as mentioned in Sec.~\ref{sec:assumptions_effect}. The extent of such quasar near-zone in the 21~cm forest spectra is sensitive to the integrated lifetime of the background quasar as opposed to the Ly$\alpha$ forest near-zone size, which is dictated by the latest quasar accretion episode. Therefore, the 21~cm forest in the proximity of the background radio-loud quasars can be used to test models of supermassive black hole growth \citep{Soltinsky_2023}.

Besides constraining the mechanics driving the quasars, \citet{Ewall_Wice_2014} suggest that the 21~cm forest features in the tomographic 21~cm power spectrum provide information on the high-$z$ radio-loud quasar population. More information on these sources can be found in Sec.~\ref{sec:radio_sources}.

\subsection{Cosmology}

The 21 cm forest is an essential tool for probing small-scale cosmological structures and dark matter properties. By analyzing the absorption lines produced by neutral hydrogen in minihalos, this technique enables exploration of early universe dynamics and the thermal history of the IGM. The 21 cm forest is particularly useful for studying small-scale structures that are beyond the reach of other methods such as the Lyman-$\alpha$ forest. One of its key applications is the investigation of ultralight dark matter (ULDM), including axion-like particles (ALPs), which suppress small-scale structures via quantum pressure effects. The number of 21 cm absorption lines can be used to constrain ULDM masses as low as \( 10^{-18} \) eV, far surpassing the capabilities of Lyman-$\alpha$ forest observations\citep{Shimabukuro_2020a}. Additionally, scenarios involving axion-like particles in post-inflationary Peccei–Quinn symmetry breaking can also be probed, in which isocurvature perturbations enhance small-scale structures and increase the number of 21 cm absorption lines. This allows for the detection of ALP masses ranging from \( 10^{-18} \) eV to \( 10^{-6} \) eV, depending on the temperature dependence of the axion mass\citep{Shimabukuro_2020b}.

Beyond dark matter, the 21 cm forest also provides insights into astrophysical and dynamical processes that affect structure formation. The relative velocity between dark matter and baryons, a result of different evolution rates before photon decoupling, suppresses the growth of minihalos. This suppression reduces the number of 21 cm absorption lines, and its scale dependence thereby makes it an important observable for testing early structure formation models\citep{Shimabukuro_2023}. 
Primordial black holes (PBHs) introduce shot noise isocurvature modes, which enhance the formation of minihalos and increase the number of absorption lines. However, PBHs with masses larger than \( 10 M_{\odot} \) can heat the IGM via accretion, counteracting this enhancement. This complex interplay allows the 21 cm forest to provide constraints on PBH abundances, complementing other observational techniques\citep{Villanueva-Domingo_2023}. Furthermore, the 21 cm forest can also probe the presence of long-lived scalar field configurations, such as oscillons arising from ultralight axion-like fields, which generate additional small-scale fluctuations and enhance the number of 21 cm absorption lines. By analyzing the statistical distribution of these lines, the 21 cm forest offers a way to detect non-trivial dark matter structures and distinguish them from standard cold dark matter models\citep{Kawasaki_2021}.

\subsection{Exotic physics}

In addition to astrophysical sources, exotic energy-injection processes—including dark matter annihilation \citep{Cang_2023,Nishizawa_2025}, dark matter decay\citep{Sun_2023,Facchinetti_2023}, Hawking radiation \citep{Cang_2021}, and PBH accretion \citep{Villanueva-Domingo_2023}—can inject energetic particles into the IGM.

Electromagnetic cascades of these particles can heat and ionize the IGM,
raising its temperature and ionization fraction.
As illustrated  in \cite{Vasiliev_2013,Villanueva-Domingo_2023},
additional heating and ionization from these processes tend to suppress both the optical depth and the 21~cm absorption signal.

While energy injection from astrophysical sources and decaying dark matter generally scales with the dark matter density $\rho_{\rm c}$ \citep{Sun_2023}, the annihilation rate scales as $\rho_{\rm c}^2$ and can be significantly enhanced by 
structure formation at redshifts 
targeted by the 21~cm forest \citep{Cang_2023}. Machine learning–based analyses have recently highlighted the potential of convolutional neural networks to constrain such annihilation scenarios directly from simulated 21 cm signals \citep{Nishizawa_2025}. Energy injection from PBHs may receive characteristic boosts at relevant redshifts, either through violent explosions of light PBHs ($\lesssim 10^{15}$\,g) reaching the end of their lives via Hawking evaporation \citep{Cang_2021}, or through amplified accretion luminosity of massive PBHs (with initial mass $\gtrsim 10\,M_\odot$) that undergo substantial mass growth via accretion \citep{Zhang_2025}. These characteristic boosts and evolutionary features can enhance their impact on the IGM and potentially help break degeneracies with background astrophysical processes.
Complementary to these approaches, studies of small-scale dark matter clumping \citep{Sikder_2025} show that the enhanced substructure can further amplify the annihilation signal, leaving distinctive imprints on the large-scale 21 cm power spectrum.

Conversely, interactions between dark matter and baryons can transfer heat from the IGM to the dark matter, thereby lowering the IGM temperature \citep{Munoz_2018,Barkana_2018,Flitter_2023}. Such dark matter--induced cooling has been proposed to explain the unusually deep global 21~cm signal reported by EDGES \citep{Bowman_2018}. In the context of the 21~cm forest, this cooling is expected to increase the optical depth and enhance the absorption features.



\subsection{Parameter inference}

In order to effectively utilize the 21 cm forest for investigating astrophysics, cosmology, and exotic physics, parameter estimation is a crucial step. In this section, we explore different parameter estimation techniques to accurately analyze the 21 cm forest signal, particularly in complex physical environments. These methods include Fisher matrix analysis, Bayesian inference, and deep learning, which help extract key information from the 21~cm signal\citep[e.g.][]{Acharya02.2026.SKA}.

To put different approaches on a clear footing, parameter inference methods for the 21~cm forest can be broadly grouped into
(i) explicit-likelihood approaches, which require specifying an analytic likelihood (e.g., Fisher forecasts and MCMC with an assumed likelihood for binned bandpowers),
and (ii) implicit-likelihood (likelihood-free) approaches, commonly referred to as simulation-based inference (SBI), which learn posteriors/likelihoods from forward simulations.

In this context, neural networks such as normalizing flows and U-Nets are not alternatives to SBI but are widely used building blocks within SBI frameworks: normalizing flows for density estimation, and U-Nets for data compression and feature extraction.

A common explicit-likelihood choice is a Gaussian likelihood for the (binned) power spectrum,
\begin{equation}
P(D|\boldsymbol{\theta}) \propto 
\exp\!\left[-\frac{1}{2}\sum_{k,k'}
\bigl(P_{\rm obs}(k)-P_{\rm model}(k|\boldsymbol{\theta})\bigr)\,
{\bf C}^{-1}_{kk'}\,
\bigl(P_{\rm obs}(k')-P_{\rm model}(k'|\boldsymbol{\theta})\bigr)\right],
\end{equation}
where the covariance ${\bf C}_{kk'}$ includes both thermal noise and sample (cosmic) variance.
This Gaussian, bandpower-based likelihood is an approximation whose validity depends primarily on the effective number of independent modes/sightlines and the degree of non-Gaussianity in the signal and its covariance, rather than on S/N alone; in noise-dominated regimes the likelihood often becomes more nearly Gaussian, whereas sparse sampling and non-Gaussian covariances can bias Gaussian-likelihood inference.

Studies in the 21~cm literature have shown that SBI can yield accurate posteriors with training sets smaller than the number of forward-model evaluations required for traditional MCMC \citep{Saxena_2023, Prelogovic_2023}.
For the 21~cm forest, however, generating the simulation training set can remain costly due to the required dynamic range and cross-scale modelling.
This motivates amortized pipelines that combine SBI with fast emulators for the forward model.

As an example of amortized inference using neural density estimators within SBI frameworks, \citet{Sun_2024} proposed a method that combines generative normalizing flows (GNF) with inference normalizing flows (INF) for data augmentation and parameter inference. This method first utilizes GNF to generate high-fidelity 1D power spectra based on a small number of samples, ensuring that the coefficient of determination (\(R^2\)) between the generated results corresponding to each parameter and the simulation results exceeds \(R^2 > 0.99\). Subsequently, the observed or fiducial model's \(P(k)\) is mapped to the posterior distribution through a bijection transformation by an INF: 
\begin{equation}
  Q(\boldsymbol{\theta}\mid P)
  =\pi\!\bigl[f_P^{-1}(\boldsymbol{\theta})\bigr]\,
   \bigl|\det J_{f_P^{-1}}\bigr|,
\end{equation}
achieving millisecond-level parameter inference.

\begin{figure}[htbp]
    \centering
    \includegraphics[width=1.0\hsize]{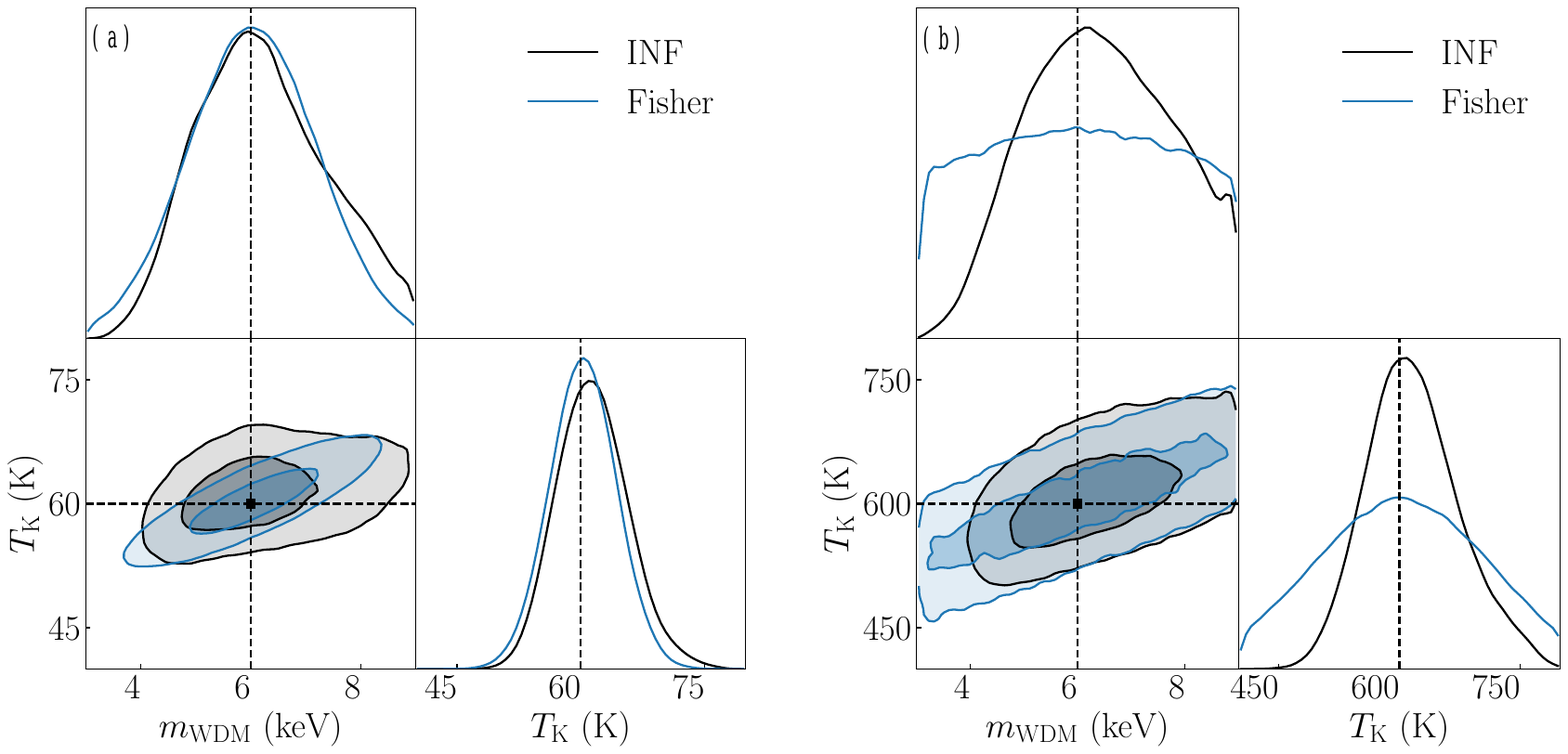}
    \caption{The posterior distributions for $m_{\mathrm{WDM}}$ and $T_K$ with the inference normalizing flow and Fisher matrix under different heating levels. The figure is taken from \citet{Sun_2024}.}
    \label{fig:pe_NF}
\end{figure}

As shown in Figure \ref{fig:pe_NF}, under low heating (\(T_K \approx 60~\mathrm{K}\)), the posterior matches the Fisher ellipse. However, in strongly heated, non-Gaussian regimes, the Fisher matrix fails to provide accurate errors, while the normalizing flow approach reliably recovers the true \(m_{\mathrm{WDM}}\) and \(T_K\). This provides an efficient and reliable analysis pipeline for the ongoing SKA 21~cm forest observations. Deep learning enhances computational efficiency, overcoming the resource demands of simulation-based methods, while maintaining excellent performance in non-Gaussian environments for robust parameter inference.

Another likelihood-free inference approach employs a U-Net architecture to denoise the 21 cm forest flux spectrum, followed by XGBoost regression for parameter inference. \citet{Patil_2026} have shown that such parameter inference analysis based on the 1D power spectra acquired from a U-Net denoised 21~cm forest outperforms traditional Bayesian methods. Moreover, omitting the power spectrum entirely and using latent features of the 21 cm forest flux spectrum yields even tighter constraints on the IGM properties and/or alleviates the requirements on the integration time of the observations.

Although deep learning and likelihood-free inference methods provide powerful tools, challenges remain in areas such as model interpretability, handling complex astrophysical signals, and reducing computational resource demands. Future research will focus on refining these techniques to improve parameter-estimation accuracy and to develop efficient, reliable analysis pipelines for ongoing SKA 21 cm forest observations.

\section{High-redshift radio sources}\label{sec:radio_sources}
\subsection{General high-z sources }

\subsubsection{Fast Radio Bursts as an independent probe}

Fast Radio Bursts (FRBs) are highly energetic transients of cosmic origin that are widely distributed across the whole sky\citep[e.g.][]{Caleb02.2026.SKA,Curtin01.2026.SKA}. The FRB continuum signal experiences dispersion while passing through the intervening ionized medium, including both Galactic and extragalactic components, which is quantified as the dispersion measure (DM). Although the origin of these events is still mysterious, the DM measurements are accurate enough to quantify the amount of plasma they pass through.

During the epoch of reionization (EoR), when the intergalactic medium (IGM) transitioned from a neutral to an ionized state, the DM measurements of FRBs could serve as a potential alternative probe of the ionized IGM, complementary to the 21 cm forest that traces the neutral hydrogen component \citep{Fialkov_2016, Yoshiura_2018}. Recent studies have found that it is possible to constrain the physical state of the IGM (characterized by the mean IGM temperature, the midpoint and width of reionization, and Thomson scattering optical depth) with only $\sim 100-1000$ FRBs observed during EoR \citep{Pagano_2021, Heimersheim_2022, Maity_2024}. Furthermore, for a given cosmology, FRBs have the potential to constrain the astrophysics of reionization \citep{Maity_2024, Shaw_2024}. 

In addition, it has been demonstrated that the sky-averaged DM and its angular variance can potentially distinguish between different reionization histories, as they are directly connected to the typical sizes of ionized bubbles. We require $\sim1000$ FRBs distributed over the entire EoR window, under ideal conditions, to place constraints on the typical bubble sizes using the variances \citep{Shaw_2024}, and reionization topologies can be further probed with an order of magnitude higher number of FRBs \citep{Ziegler_2025}.

However, the question remains regarding the detection prospects of high-redshift FRBs ($z \ge 5$). At present, highly magnetized and rapidly rotating magnetars are considered the most promising progenitors of FRBs, which are expected to arise from the deaths of massive Pop III stars. Given the increasing evidence for these massive stars during the EoR and the improved sensitivity of the upcoming SKA telescopes, there is a strong potential to detect a large number of FRBs during the EoR \citep{Fialkov_2017, Hashimoto_2020, Gupta_2025}, which would provide an ionized-gas counterpart to the neutral-gas information revealed by the 21 cm forest, together offering a more complete picture of the early Universe.


\subsubsection{Radio galaxies}
In 2018, the EDGES experiment reported the first detection of the global 21~cm differential brightness temperature \citep{Bowman_2018}, revealing an absorption trough at redshift $z \sim 17$ with a depth of $\delta T_{\rm b} \approx -500$ mK.
This signal is roughly twice as deep as that predicted by standard models, which assume a radio background dominated by the CMB. Early radio-loud galaxies capable of producing an excess radio background comparable to the CMB could potentially explain such a deep signal \citep{Feng_2018, Reis_2020, Fialkov_2019, mirocha_2019}, and consequently leave detectable imprints on 21~cm forest signals.

However, for typical atomic-cooling galaxies (ACGs) to achieve the depth observed by EDGES, the required radio background strength would exceed the current ARCADE2 upper limits \citep{Fixsen_2009} by orders of magnitude \citep{mirocha_2019, Reis_2020}. Furthermore, the corresponding star-formation rate density would also surpass the limits inferred from UV luminosity functions by nearly two orders of magnitude \citep{mirocha_2019}.

In contrast, molecular-cooling galaxies (MCGs), which reside in minihalos with masses $\lesssim 10^6M_\odot$, as opposed to ACGs that inhabit halos with masses $\gtrsim 10^8M_\odot$ \citep{Munoz_2021}, are more susceptible to feedback from Lyman–Werner radiation. This radiation photodissociates $H_2$ molecules, suppressing further star formation and thereby causing the radio emission from MCGs to decline naturally once a Lyman–Werner background is established. As a result, MCGs are considered a plausible candidate population capable of generating a radio background strong enough to explain the EDGES depth without violating ARCADE2 constraints \citep{Cang_2024}.

Nevertheless, several concerns have been raised regarding the original EDGES analysis of sky-temperature data \citep{Hills_2018, Sims_2019, Cang_2024}, including the modeling of foregrounds, calibration residuals, and the parameterization of the cosmic 21~cm signal.
Most recently, \citet{Cang_2024} re-analyzed the EDGES sky-temperature data, jointly forward-modeling the foregrounds, calibration residuals, and the cosmic signal expected from MCG radio galaxies. Their analysis found that an excess radio background is in fact decisively disfavored by the data. Moreover, a subsequent measurement by SARAS3 \citep{Bevins_2022} found no evidence for the 21~cm signal initially reported by EDGES.
Therefore, the existence of a strong 21~cm global signal—and consequently of early radio galaxies responsible for it—requires further experimental verification and more careful characterization of foregrounds and instrumental systematics.

\subsection{Background sources for 21~cm forest observations}
We now turn to bright radio-loud quasars, which are the primary background sources for 21 cm forest observations. Among the various types of high-redshift sources discussed above, bright radio-loud quasars are of particular importance for 21~cm forest observations. Detecting the 21~cm forest requires such luminous background radio sources, since the absorption features are imprinted on their spectra as the signal passes through intervening neutral hydrogen. The limited number of suitable sources has long been one of the main challenges for practical 21~cm forest detection.

The number of radio-loud quasars identified at $z \ge 5.5$ (i.e., before the completion of reionization) has recently increased. Specifically, this number has more than quadrupled since the previous edition of the SKA Science Book (2015), now exceeding 30 known sources \citep[e.g.,][]{Fan_2001,McGreer_2006,Jiang_2009,Willott_2010,Zeimann_2011,Banados_2015,Banados_2021,Banados_2023,Wang_2019,Belladitta_2020,Liu_2021,Ighina_2021,Ighina_2023,Ighina_2024,Ighina_2025,Endsley_2023,Gloudemans_2022,Gloudemans_2023,Wolf_2024}.
In addition, both the redshift and brightness ranges of radio-loud quasars have expanded. For example, the most distant radio-loud quasar has been detected at $z \sim 7$ \citep{Banados_2024}, while the brightest one shows a flux density of $110.6 \pm 13.8\,\rm mJy$ at 150\,MHz \citep{Banados_2018}.
Alternative potential background sources include gamma-ray burst afterglows \citep{Ioka_2005,Ciardi_2015} and radio galaxies \citep{Drouart_2020}.
With these newly discovered background radio sources, the prospects for detecting the 21~cm forest have significantly improved.

\begin{figure}
\centering
\includegraphics[width=0.5\hsize]{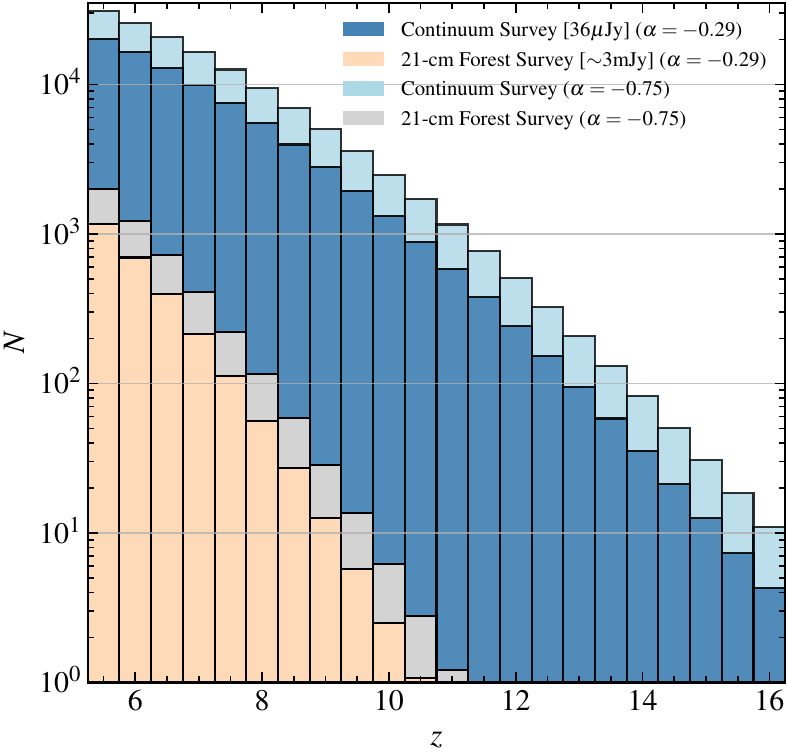}
\caption{Predicted redshift distribution of radio-loud quasars observable with one year of SKA-Low observations. The “21~cm Forest Survey” refers to high-spectral-resolution observations suitable for 21~cm forest analysis. The figure is taken from \citet{Niu_2025}.}
\label{fig:Forecast_QSO}
\end{figure}

The SKA itself also holds great promise for detecting radio-loud quasars. Recently, the Australian Square Kilometre Array Pathfinder (ASKAP) discovered six new radio-loud quasars at $z > 5.5$ through the combination of the Rapid ASKAP Continuum Survey (RACS) and deep, wide-area optical/near-infrared surveys \citep{Ighina_2025}, demonstrating the SKA’s powerful potential in this field.
Using a physically motivated model, \citet{Niu_2025} predicted the observational prospects of SKA-Low for high-$z$ radio-loud quasars. As shown in Figure~\ref{fig:Forecast_QSO}, their results indicate that a one-year sky survey with SKA-Low could detect approximately 20 radio-loud quasars at $z \sim 9$, each bright enough that, in principle, high-spectral-resolution follow-up could resolve individual 21~cm absorption lines. These quasars will serve as crucial background sources for future studies of the 21~cm forest.

It is worth noting that the discovery of new quasars relies on spectroscopic observations in the optical and near-infrared bands to determine their redshifts.
For quasars in the epoch of reionization, the Lyman-$\alpha$ break in their spectra is redshifted into the infrared due to cosmic expansion.
Meanwhile, spectral features in the near-infrared are crucial for efficiently identifying quasars among the numerous point sources detected in radio observations \citep[e.g.,][]{Liu_2021,Yang_2023,Euclid_2024}
.
Therefore, future observations of high-$z$ radio-loud quasars will require a combination of SKA and advanced infrared telescopes.
\citet{Niu_2025} assessed the observational prospects of Euclid \citep{Euclid_2022}
 and the Nancy Grace Roman Space Telescope \citep{Spergel_2015} for radio-loud quasars at $z \sim 9$.
Their results show that both space-based infrared surveys can detect the majority of radio-loud quasars at $z \sim 9$ that SKA-Low could observe with resolved individual 21~cm lines.
The synergy between SKA and space-based infrared telescopes will provide a crucial foundation for identifying radio-loud quasars that serve as background sources for 21~cm forest studies at the highest redshifts.

We may also note that JWST has detected more high-$z$ AGNs than previously expected in the pre-JWST era \citep{Harikane_2023}, and black holes with higher masses compared to those predicted by the local $M_\bullet$–$M_*$ relation \citep{Maiolino_2024}.
These black holes have masses of $\sim10^6$–$10^8,M_\odot$, indicating that SMBH seeds formed efficiently and grew rapidly during the Cosmic Dawn and EoR \citep{Goulding_2023,Jeon_2025}.
Although attempts to detect the radio emission from JWST-detected AGNs have not revealed strong signals \citep{Mazzolari_2024,Gloudemans_2025}, this does not rule out the possibility that some of them are radio-loud.
Even if only a small fraction of these high-$z$ black holes enter a radio-loud phase, they could serve as potential background sources for 21~cm forest observations.

\section{Summary \& Discussion}

The 21~cm forest, seen as neutral hydrogen (HI) absorption lines in the spectra of distant radio sources, is a unique probe of small-scale cosmic structure and the thermal history of the early universe. Recent progress has made both theoretical and observational studies increasingly feasible. Its modeling includes analytic, semi-analytic, hydrodynamical, and semi-numerical approaches: analytic methods efficiently explore large parameter spaces, while numerical simulations capture the detailed physics of the intergalactic medium (IGM). In particular, hydrodynamic simulations with radiative transfer allow sophisticated modeling of absorption features and their astrophysical dependence. Statistical tools such as the 1D power spectrum and WST are particularly valuable for data analysis. The power spectrum improves signal-to-noise and helps separate astrophysical and dark matter effects, whereas WST captures non-Gaussian structures, localized fluctuations, and cross-scale interactions. The 21~cm forest can constrain the thermal and ionization state of the IGM, supermassive black hole growth, reionization history, and cosmological physics such as ultralight dark matter, primordial black holes, and baryon-dark matter relative velocities. The SKA will greatly enhance the sensitivity of these observations, enabling major progress in early-universe studies.

We should prioritize refining observational strategies and analysis pipelines to optimize SKA data for 21~cm forest measurements. Because the forest signal appears as an extremely weak, narrow-band absorption feature on bright continuum spectra, the main near-term limitations are likely to come from spectral systematics rather than raw sensitivity alone. Key priorities therefore include controlling frequency-dependent instrumental effects—such as residual bandpass/gain ripples, beam chromaticity, polarization leakage, ionospheric variability, and spectral structure from imperfect RFI flagging—through improved calibration, robust quality assessment, and end-to-end validation, including signal-injection tests, null tests, and independent pipeline cross-checks. Accurate modeling of the intrinsic spectra of background radio sources is also essential to distinguish genuine absorption from calibration-induced structure. Given these limitations, AA* and science-verification observations will likely first provide robust constraints from large-scale spectral statistics, such as the mean optical depth, the IGM spin temperature, and the ionization history, before enabling detailed individual-line studies and tighter model discrimination.

In parallel, a statistical sample of high-redshift FRBs offers an independent probe of the IGM evolution and morphology, complementing 21~cm observations. With the sensitivity of the SKA, detecting such events during the reionization epoch may become feasible. This will require close integration with multi-wavelength observations, especially from JWST, Euclid, and the Nancy Grace Roman Space Telescope. Combining SKA radio data with deep optical/IR imaging and spectroscopy will allow robust identification, redshift measurement, and physical characterization of high-redshift radio-loud quasars, galaxies, and FRB hosts. Such synergy is crucial for constructing reliable background-source samples for 21~cm forest studies and for cross-validating interpretations of absorption features. Optical/IR data constrain the ionization state of the IGM and reionization history, while the 21~cm forest probes small-scale structure and thermal conditions. Future work should focus on coordinated surveys, joint analysis pipelines, and statistical frameworks for multi-wavelength cross-correlation.

Finally, expanding parameter-estimation methods with Bayesian inference and deep-learning-based likelihood-free approaches will greatly improve both the accuracy and efficiency of data analysis. Traditional methods such as Fisher analysis and MCMC remain powerful, but are limited by Gaussian assumptions and computational cost. Future work should therefore emphasize flexible Bayesian frameworks that can handle non-Gaussian, multi-modal, and otherwise complex likelihoods, especially in regimes of strong heating or low signal-to-noise. At the same time, deep-learning approaches enable rapid posterior estimation in high-dimensional spaces without requiring explicit likelihoods. Integrating these methods into analysis pipelines will allow more robust extraction of astrophysical and cosmological information from SKA data, helping to fully realize the potential of the 21~cm forest as a probe of the early Universe.

\section{Author contributions}

H.S. and Y.X. coordinated this chapter. H.S. contributed to Secs. 1, 3.1, 3.2, 4.2, 4.4, and 6. T.Š. contributed to Secs. 1, 2.1, 2.2, 3.1, 3.2, 4.1, 4.4, and 5.2. Y.X. contributed to the abstract and Secs. 1, 2, 3.2, and 4.4. T.-Y.S. and Y.S. contributed to Secs. 3.2 and 4.4. Q.N. and B.Y. contributed to Sec. 5.2. X.Z. contributed to Secs. 3.2, 4.4, and 5.2. T.T.T. and K.Y. contributed to Sec. 3.2. A.K.S. and B.M. contributed to Sec. 5.1. J.C. contributed to Secs. 4.3 and 5.1.2. A.J.N. contributed to Sec. 4.3. B.C. contributed to Secs. 2.1 and 2.2.

\section*{Acknowledgement}
HS is supported by the National SKA Program of China (No. 2020SKA0110401), the NSFC (Grant No.~12103044), and the Yunnan Provincial Key Laboratory of Survey Science (Project No. 202449CE340002).
TŠ acknowledges support from the Istituto Nazionale di Astrofisica Osservatorio Astronomico di Trieste (INAF-OATs) through the Theory grant ``Cosmological Investigation of the Cosmic Web'' (C93C23006820005), and from the Istituto Nazionale di Fisica Nucleare (INFN) INDARK grant.
YX acknowledges support from the National SKA Program of China (No. 2020SKA0110401) and the National Key R\&D Program of China (No. 2022YFF0504300).
BY acknowledges support from the National SKA Program of China (No. 2020SKA0110402) and the NSFC International (Regional) Cooperation and Exchange Project (No. 12361141814).
TTT acknowledges JSPS Grants-in-Aid for Scientific Research (21H01128 and 24H00247), and partial support from the ISM Collaboration Funding ``Machine-Learning-Based Cosmogony: From Structure Formation to Galaxy Evolution''.
KY is supported by JSPS KAKENHI (JP21H01079, JP25H00625) and by MEXT’s ``Program for Promoting Researches on the Supercomputer Fugaku'' (Structure and Evolution of the Universe Unraveled by Fusion of Simulation and AI).
XZ is supported by the National SKA Program of China (2022SKA0110200, 2022SKA0110203), the National Natural Science Foundation of China (12473091, 12473001, 12533001), and the National 111 Project (B16009).
AJN is supported by JSPS Grant-in-Aid for Transformative Research Areas (JP25H01551) and JSPS International Leading Research (22K21349).
AKS acknowledges support from the National Science Foundation (Grant No. 2206602).





\bibliographystyle{abbrvnat-maxbibnames4}

\bibliography{chapter} 

\begin{thebibliography}{165}
\providecommand{\natexlab}[1]{#1}
\providecommand{\url}[1]{\texttt{#1}}
\expandafter\ifx\csname urlstyle\endcsname\relax
  \providecommand{\doi}[1]{doi: #1}\else
  \providecommand{\doi}{doi: \begingroup \urlstyle{rm}\Url}\fi

\bibitem[Abdurashidova et~al.(2022)]{Abdurashidova_2022}
Z.~Abdurashidova et~al.
\newblock \emph{The Astrophysical Journal}, 925\penalty0 (2):\penalty0 221, 2022.
\newblock \doi{10.3847/1538-4357/ac1c78}.

\bibitem[Abdurashidova et~al.(2023)]{Abdurashidova_2023}
Z.~Abdurashidova et~al.
\newblock \emph{The Astrophysical Journal}, 945\penalty0 (2):\penalty0 124, 2023.
\newblock \doi{10.3847/1538-4357/acaf50}.

\bibitem[Acharya et~al.(2026{\natexlab{a}})Acharya, author2, author3, author4, and author5]{Acharya01.2026.SKA}
A.~Acharya et al.
\newblock In \emph{Advancing Astrophysics with the SKA -- II (AASKAII)}. 2026{\natexlab{a}}.
\newblock arXiv search: Report number AASKAII/Acharya01.

\bibitem[Acharya et~al.(2026{\natexlab{b}})Acharya, author2, author3, author4, and author5]{Acharya02.2026.SKA}
A.~Acharya et al.
\newblock In \emph{Advancing Astrophysics with the SKA -- II (AASKAII)}. 2026{\natexlab{b}}.
\newblock arXiv search: Report number AASKAII/Acharya02.

\bibitem[{Ba{\~n}ados} et~al.(2015){Ba{\~n}ados}, {Venemans}, {Morganson}, {Hodge}, {Decarli}, {Walter}, {Stern}, {Schlafly}, {Farina}, {Greiner}, {Chambers}, {Fan}, {Rix}, {Burgett}, {Draper}, {Flewelling}, {Kaiser}, {Metcalfe}, {Morgan}, {Tonry}, and {Wainscoat}]{Banados_2015}
E.~{Ba{\~n}ados} et al.
\newblock \emph{\apj}, 804\penalty0 (2):\penalty0 118, May 2015.
\newblock \doi{10.1088/0004-637X/804/2/118}.

\bibitem[{Ba{\~n}ados} et~al.(2018){Ba{\~n}ados}, {Carilli}, {Walter}, {Momjian}, {Decarli}, {Farina}, {Mazzucchelli}, and {Venemans}]{Banados_2018}
E.~{Ba{\~n}ados} et al.
\newblock \emph{\apjl}, 861\penalty0 (2):\penalty0 L14, July 2018.
\newblock \doi{10.3847/2041-8213/aac511}.

\bibitem[{Ba{\~n}ados} et~al.(2021){Ba{\~n}ados}, {Mazzucchelli}, {Momjian}, {Eilers}, {Wang}, {Schindler}, {Connor}, {Andika}, {Barth}, {Carilli}, {Davies}, {Decarli}, {Fan}, {Farina}, {Hennawi}, {Pensabene}, {Stern}, {Venemans}, {Wenzl}, and {Yang}]{Banados_2021}
E.~{Ba{\~n}ados} et al.
\newblock \emph{\apj}, 909\penalty0 (1):\penalty0 80, Mar. 2021.
\newblock \doi{10.3847/1538-4357/abe239}.

\bibitem[{Ba{\~n}ados} et~al.(2023){Ba{\~n}ados}, {Schindler}, {Venemans}, {Connor}, {Decarli}, {Farina}, {Mazzucchelli}, {Meyer}, {Stern}, {Walter}, {Fan}, {Hennawi}, {Khusanova}, {Morrell}, {Nanni}, {Noirot}, {Pensabene}, {Rix}, {Simon}, {Verdoes Kleijn}, {Xie}, {Yang}, and {Connor}]{Banados_2023}
E.~{Ba{\~n}ados} et al.
\newblock \emph{\apjs}, 265\penalty0 (1):\penalty0 29, Mar. 2023.
\newblock \doi{10.3847/1538-4365/acb3c7}.

\bibitem[{Ba{\~n}ados} et~al.(2024){Ba{\~n}ados}, {Momjian}, {Connor}, {Belladitta}, {Decarli}, {Mazzucchelli}, {Venemans}, {Walter}, {Wang}, {Xie}, {Barth}, {Eilers}, {Fan}, {Khusanova}, {Schindler}, {Stern}, {Yang}, {Andika}, {Carilli}, {Farina}, {Fabian}, {Hennawi}, {Pensabene}, and {Rojas-Ruiz}]{Banados_2024}
E.~{Ba{\~n}ados} et al.
\newblock \emph{Nature Astronomy}, Dec. 2024.
\newblock \doi{10.1038/s41550-024-02431-4}.

\bibitem[Bag et~al.(2026)Bag, author2, author3, author4, and author5]{Bag01.2026.SKA}
S.~Bag et al.
\newblock In \emph{Advancing Astrophysics with the SKA -- II (AASKAII)}. 2026.
\newblock arXiv search: Report number AASKAII/Bag01.

\bibitem[Barkana(2018)]{Barkana_2018}
R.~Barkana.
\newblock \emph{Nature}, 555\penalty0 (7694):\penalty0 71--74, 2018.
\newblock \doi{10.1038/nature25791}.

\bibitem[Barkana et~al.(2026)Barkana, author2, author3, author4, and author5]{Barkana01.2026.SKA}
R.~Barkana et al.
\newblock In \emph{Advancing Astrophysics with the SKA -- II (AASKAII)}. 2026.
\newblock arXiv search: Report number AASKAII/Barkana01.

\bibitem[Beardsley et~al.(2016)]{Beardsley_2016}
A.~P. Beardsley et~al.
\newblock \emph{The Astrophysical Journal}, 833\penalty0 (1):\penalty0 102, 2016.
\newblock \doi{10.3847/1538-4357/833/1/102}.

\bibitem[{Becker} et~al.(2015){Becker}, {Bolton}, {Madau}, {Pettini}, {Ryan-Weber}, and {Venemans}]{Becker_2015}
G.~D. {Becker} et al.
\newblock \emph{\mnras}, 447\penalty0 (4):\penalty0 3402--3419, Mar. 2015.
\newblock \doi{10.1093/mnras/stu2646}.

\bibitem[{Becker} et~al.(2021){Becker}, {D'Aloisio}, {Christenson}, {Zhu}, {Worseck}, and {Bolton}]{Becker_2021}
G.~D. {Becker} et al.
\newblock \emph{\mnras}, 508\penalty0 (2):\penalty0 1853--1869, Dec. 2021.
\newblock \doi{10.1093/mnras/stab2696}.

\bibitem[{Becker} et~al.(2024){Becker}, {Bolton}, {Zhu}, and {Hashemi}]{Becker_2024}
G.~D. {Becker}, J.~S. {Bolton}, Y.~{Zhu}, and S.~{Hashemi}.
\newblock \emph{\mnras}, 533\penalty0 (2):\penalty0 1525--1540, Sept. 2024.
\newblock \doi{10.1093/mnras/stae1918}.

\bibitem[{Belladitta} et~al.(2020){Belladitta}, {Moretti}, {Caccianiga}, {Spingola}, {Severgnini}, {Della Ceca}, {Ghisellini}, {Dallacasa}, {Sbarrato}, {Cicone}, {Cassar{\`a}}, and {Pedani}]{Belladitta_2020}
S.~{Belladitta} et al.
\newblock \emph{\aap}, 635:\penalty0 L7, Mar. 2020.
\newblock \doi{10.1051/0004-6361/201937395}.

\bibitem[Bevins et~al.(2022)Bevins, Fialkov, Acedo, Handley, Singh, Subrahmanyan, and Barkana]{Bevins_2022}
H.~T.~J. Bevins et al.
\newblock \emph{Nature Astron.}, 6\penalty0 (12):\penalty0 1473--1483, 2022.
\newblock \doi{10.1038/s41550-022-01825-6}.

\bibitem[{Bhagwat} et~al.(2022){Bhagwat}, {Ciardi}, {Zackrisson}, and {Schaye}]{Bhagwat_2022}
A.~{Bhagwat}, B.~{Ciardi}, E.~{Zackrisson}, and J.~{Schaye}.
\newblock \emph{\mnras}, 517\penalty0 (2):\penalty0 2331--2342, Dec. 2022.
\newblock \doi{10.1093/mnras/stac2663}.

\bibitem[{Bosman} et~al.(2022){Bosman}, {Davies}, {Becker}, {Keating}, {Davies}, {Zhu}, {Eilers}, {D'Odorico}, {Bian}, {Bischetti}, {Cristiani}, {Fan}, {Farina}, {Haehnelt}, {Hennawi}, {Kulkarni}, {Mesinger}, {Meyer}, {Onoue}, {Pallottini}, {Qin}, {Ryan-Weber}, {Schindler}, {Walter}, {Wang}, and {Yang}]{Bosman_2022}
S.~E.~I. {Bosman} et al.
\newblock \emph{\mnras}, 514\penalty0 (1):\penalty0 55--76, July 2022.
\newblock \doi{10.1093/mnras/stac1046}.

\bibitem[Bowman et~al.(2018)Bowman, Rogers, Monsalve, Mozdzen, and Mahesh]{Bowman_2018}
J.~D. Bowman et al.
\newblock \emph{Nature}, 555\penalty0 (7694):\penalty0 67--70, 2018.
\newblock \doi{10.1038/nature25792}.

\bibitem[Burba et~al.(2026)Burba, author2, author3, author4, and author5]{Burba01.2026.SKA}
J.~Burba et al.
\newblock In \emph{Advancing Astrophysics with the SKA -- II (AASKAII)}. 2026.
\newblock arXiv search: Report number AASKAII/Burba01.

\bibitem[{Cain} et~al.(2021){Cain}, {D'Aloisio}, {Gangolli}, and {Becker}]{Cain_2021}
C.~{Cain}, A.~{D'Aloisio}, N.~{Gangolli}, and G.~D. {Becker}.
\newblock \emph{\apjl}, 917\penalty0 (2):\penalty0 L37, Aug. 2021.
\newblock \doi{10.3847/2041-8213/ac1ace}.

\bibitem[Caleb et~al.(2026)Caleb, author2, author3, author4, and author5]{Caleb02.2026.SKA}
M.~Caleb et al.
\newblock In \emph{Advancing Astrophysics with the SKA -- II (AASKAII)}. 2026.
\newblock arXiv search: Report number AASKAII/Caleb02.

\bibitem[Cang et~al.(2022)Cang, Gao, and Ma]{Cang_2021}
J.~Cang, Y.~Gao, and Y.-Z. Ma.
\newblock \emph{JCAP}, 03\penalty0 (03):\penalty0 012, 2022.
\newblock \doi{10.1088/1475-7516/2022/03/012}.

\bibitem[{Cang} et~al.(2025){Cang}, {Gao}, and {Ma}]{Cang_2023}
J.~{Cang}, Y.~{Gao}, and Y.-Z. {Ma}.
\newblock \emph{\prd}, 112\penalty0 (10):\penalty0 103048, Nov. 2025.
\newblock \doi{10.1103/69jk-m5xg}.

\bibitem[Cang et~al.(2025)Cang, Mesinger, Murray, Breitman, Qin, and Trotta]{Cang_2024}
J.~Cang et al.
\newblock \emph{Astron. Astrophys.}, 698:\penalty0 A152, 2025.
\newblock \doi{10.1051/0004-6361/202452982}.

\bibitem[{Carilli} et~al.(2002){Carilli}, {Gnedin}, and {Owen}]{Carilli_2002}
C.~L. {Carilli}, N.~Y. {Gnedin}, and F.~{Owen}.
\newblock \emph{\apj}, 577\penalty0 (1):\penalty0 22--30, Sept. 2002.
\newblock \doi{10.1086/342179}.

\bibitem[{Chan} et~al.(2024){Chan}, {Ben{\'\i}tez-Llambay}, {Theuns}, {Frenk}, and {Bower}]{Chan_2024}
T.~K. {Chan} et al.
\newblock \emph{\mnras}, 528\penalty0 (2):\penalty0 1296--1326, Feb. 2024.
\newblock \doi{10.1093/mnras/stae114}.

\bibitem[{Chardin} et~al.(2017){Chardin}, {Puchwein}, and {Haehnelt}]{Chardin_2017}
J.~{Chardin}, E.~{Puchwein}, and M.~G. {Haehnelt}.
\newblock \emph{\mnras}, 465\penalty0 (3):\penalty0 3429--3445, Mar. 2017.
\newblock \doi{10.1093/mnras/stw2943}.

\bibitem[{Choudhury} et~al.(2021){Choudhury}, {Paranjape}, and {Bosman}]{Choudhury_2021}
T.~R. {Choudhury}, A.~{Paranjape}, and S.~E.~I. {Bosman}.
\newblock \emph{\mnras}, 501\penalty0 (4):\penalty0 5782--5796, Mar. 2021.
\newblock \doi{10.1093/mnras/stab045}.

\bibitem[{Christenson} et~al.(2021){Christenson}, {Becker}, {Furlanetto}, {Davies}, {Malkan}, {Zhu}, {Boera}, and {Trapp}]{Christenson_2021}
H.~M. {Christenson} et al.
\newblock \emph{\apj}, 923\penalty0 (1):\penalty0 87, Dec. 2021.
\newblock \doi{10.3847/1538-4357/ac2a34}.

\bibitem[{Ciardi} et~al.(2006){Ciardi}, {Scannapieco}, {Stoehr}, {Ferrara}, {Iliev}, and {Shapiro}]{Ciardi_2006}
B.~{Ciardi} et al.
\newblock \emph{\mnras}, 366\penalty0 (2):\penalty0 689--696, Feb. 2006.
\newblock \doi{10.1111/j.1365-2966.2005.09908.x}.

\bibitem[{Ciardi} et~al.(2012){Ciardi}, {Bolton}, {Maselli}, and {Graziani}]{Ciardi_2012}
B.~{Ciardi}, J.~S. {Bolton}, A.~{Maselli}, and L.~{Graziani}.
\newblock \emph{\mnras}, 423\penalty0 (1):\penalty0 558--574, June 2012.
\newblock \doi{10.1111/j.1365-2966.2012.20902.x}.

\bibitem[{Ciardi} et~al.(2013){Ciardi}, {Labropoulos}, {Maselli}, {Thomas}, {Zaroubi}, {Graziani}, {Bolton}, {Bernardi}, {Brentjens}, {de Bruyn}, {Daiboo}, {Harker}, {Jelic}, {Kazemi}, {Koopmans}, {Martinez}, {Mellema}, {Offringa}, {Pandey}, {Schaye}, {Veligatla}, {Vedantham}, and {Yatawatta}]{Ciardi_2013}
B.~{Ciardi} et al.
\newblock \emph{\mnras}, 428\penalty0 (2):\penalty0 1755--1765, Jan. 2013.
\newblock \doi{10.1093/mnras/sts156}.

\bibitem[{Ciardi} et~al.(2015){Ciardi}, {Inoue}, {Abdalla}, {Asad}, {Bernardi}, {Bolton}, {Brentjens}, {de Bruyn}, {Chapman}, {Daiboo}, {Fernand ez}, {Ghosh}, {Graziani}, {Harker}, {Iliev}, {Jeli{\'c}}, {Jensen}, {Kazemi}, {Koopmans}, {Martinez}, {Maselli}, {Mellema}, {Offringa}, {Pandey}, {Schaye}, {Thomas}, {Vedantham}, {Yatawatta}, and {Zaroubi}]{Ciardi_2015}
B.~{Ciardi} et al.
\newblock \emph{\mnras}, 453\penalty0 (1):\penalty0 101--105, Oct. 2015.
\newblock \doi{10.1093/mnras/stv1640}.

\bibitem[Curtin et~al.(2026)Curtin, author2, author3, author4, and author5]{Curtin01.2026.SKA}
A.~P. Curtin et al.
\newblock In \emph{Advancing Astrophysics with the SKA -- II (AASKAII)}. 2026.
\newblock arXiv search: Report number AASKAII/Curtin01.

\bibitem[{D'Aloisio} et~al.(2015){D'Aloisio}, {McQuinn}, and {Trac}]{DAloisio_2015}
A.~{D'Aloisio}, M.~{McQuinn}, and H.~{Trac}.
\newblock \emph{\apjl}, 813\penalty0 (2):\penalty0 L38, Nov. 2015.
\newblock \doi{10.1088/2041-8205/813/2/L38}.

\bibitem[{Davies} and {Furlanetto}(2016)]{Davies_2016}
F.~B. {Davies} and S.~R. {Furlanetto}.
\newblock \emph{\mnras}, 460\penalty0 (2):\penalty0 1328--1339, Aug. 2016.
\newblock \doi{10.1093/mnras/stw931}.

\bibitem[de~Lera~Acedo et~al.(2026)de~Lera~Acedo, author2, author3, author4, and author5]{deLeraAcedo01.2026.SKA}
E.~de~Lera~Acedo et al.
\newblock In \emph{Advancing Astrophysics with the SKA -- II (AASKAII)}. 2026.
\newblock arXiv search: Report number AASKAII/deLeraAcedo01.

\bibitem[{Drouart} et~al.(2020){Drouart}, {Seymour}, {Galvin}, {Afonso}, {Callingham}, {De Breuck}, {Johnston-Hollitt}, {Kapi{\'n}ska}, {Lehnert}, and {Vernet}]{Drouart_2020}
G.~{Drouart} et al.
\newblock \emph{\pasa}, 37:\penalty0 e026, July 2020.
\newblock \doi{10.1017/pasa.2020.6}.

\bibitem[{Endsley} et~al.(2023){Endsley}, {Stark}, {Lyu}, {Wang}, {Yang}, {Fan}, {Smit}, {Bouwens}, {Hainline}, and {Schouws}]{Endsley_2023}
R.~{Endsley} et al.
\newblock \emph{\mnras}, 520\penalty0 (3):\penalty0 4609--4620, Apr. 2023.
\newblock \doi{10.1093/mnras/stad266}.

\bibitem[{Euclid Collaboration} et~al.(2022){Euclid Collaboration}, {Scaramella}, {Amiaux}, {Mellier}, {Burigana}, {Carvalho}, {Cuillandre}, {Da Silva}, {Derosa}, {Dinis}, {Maiorano}, {Maris}, {Tereno}, {Laureijs}, {Boenke}, {Buenadicha}, {Dupac}, {Gaspar Venancio}, {G{\'o}mez-{\'A}lvarez}, {Hoar}, {Lorenzo Alvarez}, {Racca}, {Saavedra-Criado}, {Schwartz}, {Vavrek}, {Schirmer}, {Aussel}, {Azzollini}, {Cardone}, {Cropper}, {Ealet}, {Garilli}, {Gillard}, {Granett}, {Guzzo}, {Hoekstra}, {Jahnke}, {Kitching}, {Maciaszek}, {Meneghetti}, {Miller}, {Nakajima}, {Niemi}, {Pasian}, {Percival}, {Pottinger}, {Sauvage}, {Scodeggio}, {Wachter}, {Zacchei}, {Aghanim}, {Amara}, {Auphan}, {Auricchio}, {Awan}, {Balestra}, {Bender}, {Bodendorf}, {Bonino}, {Branchini}, {Brau-Nogue}, {Brescia}, {Candini}, {Capobianco}, {Carbone}, {Carlberg}, {Carretero}, {Casas}, {Castander}, {Castellano}, {Cavuoti}, {Cimatti}, {Cledassou}, {Congedo}, {Conselice}, {Conversi}, {Copin}, {Corcione}, {Costille}, {Courbin}, {Degaudenzi}, {Douspis},
  {Dubath}, {Duncan}, {Dusini}, {Farrens}, {Ferriol}, {Fosalba}, {Fourmanoit}, {Frailis}, {Franceschi}, {Franzetti}, {Fumana}, {Gillis}, {Giocoli}, {Grazian}, {Grupp}, {Haugan}, {Holmes}, {Hormuth}, {Hudelot}, {Kermiche}, {Kiessling}, {Kilbinger}, {Kohley}, {Kubik}, {K{\"u}mmel}, {Kunz}, {Kurki-Suonio}, {Lahav}, {Ligori}, {Lilje}, {Lloro}, {Mansutti}, {Marggraf}, {Markovic}, {Marulli}, {Massey}, {Maurogordato}, {Melchior}, {Merlin}, {Meylan}, {Mohr}, {Moresco}, {Morin}, {Moscardini}, {Munari}, {Nichol}, {Padilla}, {Paltani}, {Peacock}, {Pedersen}, {Pettorino}, {Pires}, {Poncet}, {Popa}, {Pozzetti}, {Raison}, {Rebolo}, {Rhodes}, {Rix}, {Roncarelli}, {Rossetti}, {Saglia}, {Schneider}, {Schrabback}, {Secroun}, {Seidel}, {Serrano}, {Sirignano}, {Sirri}, {Skottfelt}, {Stanco}, {Starck}, {Tallada-Cresp{\'\i}}, {Tavagnacco}, {Taylor}, {Teplitz}, {Toledo-Moreo}, {Torradeflot}, {Trifoglio}, {Valentijn}, {Valenziano}, {Verdoes Kleijn}, {Wang}, {Welikala}, {Weller}, {Wetzstein}, {Zamorani}, {Zoubian}, {Andreon},
  {Baldi}, {Bardelli}, {Boucaud}, {Camera}, {Di Ferdinando}, {Fabbian}, {Farinelli}, {Galeotta}, {Graci{\'a}-Carpio}, {Maino}, {Medinaceli}, {Mei}, {Neissner}, {Polenta}, {Renzi}, {Romelli}, {Rosset}, {Sureau}, {Tenti}, {Vassallo}, {Zucca}, {Baccigalupi}, {Balaguera-Antol{\'\i}nez}, {Battaglia}, {Biviano}, {Borgani}, {Bozzo}, {Cabanac}, and {Cappi}]{Euclid_2022}
{Euclid Collaboration} et al.
\newblock \emph{\aap}, 662:\penalty0 A112, June 2022.
\newblock \doi{10.1051/0004-6361/202141938}.

\bibitem[{Euclid Collaboration} et~al.(2024){Euclid Collaboration}, {Bisigello}, {Massimo}, {Tortora}, {Fotopoulou}, {Allevato}, {Bolzonella}, {Gruppioni}, {Pozzetti}, {Rodighiero}, {Serjeant}, {Cunha}, {Gabarra}, {Feltre}, {Humphrey}, {La Franca}, {Landt}, {Mannucci}, {Prandoni}, {Radovich}, {Ricci}, {Salvato}, {Shankar}, {Stern}, {Spinoglio}, {Vergani}, {Vignali}, {Zamorani}, {Yung}, {Charlot}, {Aghanim}, {Amara}, {Andreon}, {Auricchio}, {Baldi}, {Bardelli}, {Battaglia}, {Bender}, {Bonino}, {Branchini}, {Brau-Nogue}, {Brescia}, {Camera}, {Capobianco}, {Carbone}, {Carretero}, {Casas}, {Castander}, {Castellano}, {Cavuoti}, {Cimatti}, {Congedo}, {Conselice}, {Conversi}, {Copin}, {Corcione}, {Courbin}, {Courtois}, {Cropper}, {Da Silva}, {Degaudenzi}, {Di Giorgio}, {Dinis}, {Dupac}, {Dusini}, {Ealet}, {Farina}, {Farrens}, {Ferriol}, {Frailis}, {Franceschi}, {Franzetti}, {Fumana}, {Galeotta}, {Garilli}, {Gillis}, {Giocoli}, {Granett}, {Grazian}, {Grupp}, {Guzzo}, {Haugan}, {Holmes}, {Hook}, {Hormuth},
  {Hornstrup}, {Jahnke}, {Keih{\"a}nen}, {Kermiche}, {Kiessling}, {Kilbinger}, {Kitching}, {K{\"u}mmel}, {Kunz}, {Kurki-Suonio}, {Ligori}, {Lilje}, {Lindholm}, {Lloro}, {Maiorano}, {Mansutti}, {Marggraf}, {Markovic}, {Martinet}, {Marulli}, {Massey}, {Maurogordato}, {Medinaceli}, {Mei}, {Mellier}, {Meneghetti}, {Merlin}, {Meylan}, {Moresco}, {Moscardini}, {Munari}, {Niemi}, {Padilla}, {Paltani}, {Pasian}, {Pedersen}, {Percival}, {Pettorino}, {Polenta}, {Poncet}, {Raison}, {Rebolo}, {Renzi}, {Rhodes}, {Riccio}, {Romelli}, {Roncarelli}, {Rossetti}, {Saglia}, {Sapone}, {Sartoris}, {Schirmer}, {Schneider}, {Schrabback}, {Secroun}, {Seidel}, {Serrano}, {Sirignano}, {Sirri}, {Stanco}, {Surace}, {Tallada-Cresp{\'\i}}, {Taylor}, {Tereno}, {Toledo-Moreo}, {Torradeflot}, {Tutusaus}, {Valentijn}, {Valenziano}, {Vassallo}, {Wang}, {Zoubian}, {Zucca}, {Biviano}, {Bozzo}, {Colodro-Conde}, {Di Ferdinando}, {Fabbian}, {Graci{\'a}-Carpio}, {Marcin}, {Mauri}, {Sakr}, {Scottez}, {Tenti}, {Akrami}, {Baccigalupi}, {Ballardini},
  {Bethermin}, {Blanchard}, {Borgani}, {Borlaff}, {Bruton}, {Burigana}, {Cabanac}, {Calabro}, {Cappi}, {Carvalho}, {Castignani}, {Castro}, {Chambers}, {Cooray}, {Coupon}, {Cucciati}, {Davini}, {De Lucia}, {Desprez}, {D{\'\i}az-S{\'a}nchez}, {Di Domizio}, {Dole}, {Escartin Vigo}, {Escoffier}, {Ferrero}, {Finelli}, {Ganga}, and {Garc{\'\i}a-Bellido}]{Euclid_2024}
{Euclid Collaboration} et al.
\newblock \emph{\aap}, 691:\penalty0 A1, Nov. 2024.
\newblock \doi{10.1051/0004-6361/202450446}.

\bibitem[{Ewall-Wice} et~al.(2014){Ewall-Wice}, {Dillon}, {Mesinger}, and {Hewitt}]{Ewall_Wice_2014}
A.~{Ewall-Wice}, J.~S. {Dillon}, A.~{Mesinger}, and J.~{Hewitt}.
\newblock \emph{\mnras}, 441\penalty0 (3):\penalty0 2476--2496, July 2014.
\newblock \doi{10.1093/mnras/stu666}.

\bibitem[Ewall-Wice et~al.(2016)]{EwallWice_2016}
A.~Ewall-Wice et~al.
\newblock \emph{Monthly Notices of the Royal Astronomical Society}, 460\penalty0 (4):\penalty0 4320--4347, 2016.
\newblock \doi{10.1093/mnras/stw1022}.

\bibitem[Facchinetti et~al.(2024)Facchinetti, Lopez-Honorez, Qin, and Mesinger]{Facchinetti_2023}
G.~Facchinetti, L.~Lopez-Honorez, Y.~Qin, and A.~Mesinger.
\newblock \emph{JCAP}, 01:\penalty0 005, 2024.
\newblock \doi{10.1088/1475-7516/2024/01/005}.

\bibitem[{Fan} et~al.(2001){Fan}, {Narayanan}, {Lupton}, {Strauss}, {Knapp}, {Becker}, {White}, {Pentericci}, {Leggett}, {Haiman}, {Gunn}, {Ivezi{\'c}}, {Schneider}, {Anderson}, {Brinkmann}, {Bahcall}, {Connolly}, {Csabai}, {Doi}, {Fukugita}, {Geballe}, {Grebel}, {Harbeck}, {Hennessy}, {Lamb}, {Miknaitis}, {Munn}, {Nichol}, {Okamura}, {Pier}, {Prada}, {Richards}, {Szalay}, and {York}]{Fan_2001}
X.~{Fan} et al.
\newblock \emph{\aj}, 122\penalty0 (6):\penalty0 2833--2849, Dec. 2001.
\newblock \doi{10.1086/324111}.

\bibitem[{Feng} and {Holder}(2018)]{Feng_2018}
C.~{Feng} and G.~{Holder}.
\newblock \emph{\apjl}, 858\penalty0 (2):\penalty0 L17, May 2018.
\newblock \doi{10.3847/2041-8213/aac0fe}.

\bibitem[{Fialkov} and {Barkana}(2019)]{Fialkov_2019}
A.~{Fialkov} and R.~{Barkana}.
\newblock \emph{\mnras}, 486\penalty0 (2):\penalty0 1763--1773, June 2019.
\newblock \doi{10.1093/mnras/stz873}.

\bibitem[{Fialkov} and {Loeb}(2016)]{Fialkov_2016}
A.~{Fialkov} and A.~{Loeb}.
\newblock \emph{\jcap}, 2016\penalty0 (5):\penalty0 004, May 2016.
\newblock \doi{10.1088/1475-7516/2016/05/004}.

\bibitem[{Fialkov} and {Loeb}(2017)]{Fialkov_2017}
A.~{Fialkov} and A.~{Loeb}.
\newblock \emph{\apjl}, 846\penalty0 (2):\penalty0 L27, Sept. 2017.
\newblock \doi{10.3847/2041-8213/aa8905}.

\bibitem[{Field}(1958)]{Field_1958}
G.~B. {Field}.
\newblock \emph{Proceedings of the IRE}, 46:\penalty0 240--250, Jan. 1958.
\newblock \doi{10.1109/JRPROC.1958.286741}.

\bibitem[{Fixsen} et~al.(2011){Fixsen}, {Kogut}, {Levin}, {Limon}, {Lubin}, {Mirel}, {Seiffert}, {Singal}, {Wollack}, {Villela}, and {Wuensche}]{Fixsen_2009}
D.~J. {Fixsen} et al.
\newblock \emph{\apj}, 734\penalty0 (1):\penalty0 5, June 2011.
\newblock \doi{10.1088/0004-637X/734/1/5}.

\bibitem[Flitter and Kovetz(2024)]{Flitter_2023}
J.~Flitter and E.~D. Kovetz.
\newblock \emph{Phys. Rev. D}, 109\penalty0 (4):\penalty0 043512, 2024.
\newblock \doi{10.1103/PhysRevD.109.043512}.

\bibitem[Furlanetto(2006)]{Furlanetto_2006}
S.~R. Furlanetto.
\newblock \emph{\mnras}, 370\penalty0 (4):\penalty0 1867, Jun 2006.
\newblock ISSN 0035-8711.
\newblock \doi{10.1111/j.1365-2966.2006.10603.x}.

\bibitem[{Furlanetto}(2006)]{Furlanetto_2006b}
S.~R. {Furlanetto}.
\newblock \emph{\mnras}, 371\penalty0 (2):\penalty0 867--878, Sept. 2006.
\newblock \doi{10.1111/j.1365-2966.2006.10725.x}.

\bibitem[{Furlanetto} and {Loeb}(2002)]{Furlanetto_2002}
S.~R. {Furlanetto} and A.~{Loeb}.
\newblock \emph{\apj}, 579:\penalty0 1--9, Nov. 2002.
\newblock \doi{10.1086/342757}.

\bibitem[{Furlanetto} et~al.(2006){Furlanetto}, {Oh}, and {Briggs}]{Furlanetto_Oh_2006}
S.~R. {Furlanetto}, S.~P. {Oh}, and F.~H. {Briggs}.
\newblock \emph{\physrep}, 433\penalty0 (4-6):\penalty0 181--301, Oct. 2006.
\newblock \doi{10.1016/j.physrep.2006.08.002}.

\bibitem[{Gaikwad} et~al.(2020){Gaikwad}, {Rauch}, {Haehnelt}, {Puchwein}, {Bolton}, {Keating}, {Kulkarni}, {Ir{\v{s}}i{\v{c}}}, {Ba{\~n}ados}, {Becker}, {Boera}, {Zahedy}, {Chen}, {Carswell}, {Chardin}, and {Rorai}]{Gaikwad_2020}
P.~{Gaikwad} et al.
\newblock \emph{\mnras}, 494\penalty0 (4):\penalty0 5091--5109, June 2020.
\newblock \doi{10.1093/mnras/staa907}.

\bibitem[{Gaikwad} et~al.(2023){Gaikwad}, {Haehnelt}, {Davies}, {Bosman}, {Molaro}, {Kulkarni}, {D'Odorico}, {Becker}, {Davies}, {Nasir}, {Bolton}, {Keating}, {Ir{\v{s}}i{\v{c}}}, {Puchwein}, {Zhu}, {Asthana}, {Yang}, {Lai}, and {Eilers}]{Gaikwad_2023}
P.~{Gaikwad} et al.
\newblock \emph{\mnras}, 525\penalty0 (3):\penalty0 4093--4120, Nov. 2023.
\newblock \doi{10.1093/mnras/stad2566}.

\bibitem[{Gloudemans} et~al.(2022){Gloudemans}, {Duncan}, {Saxena}, {Harikane}, {Hill}, {Zeimann}, {R{\"o}ttgering}, {Yang}, {Best}, {Ba{\~n}ados}, {Drabent}, {Hardcastle}, {Hennawi}, {Lansbury}, {Magliocchetti}, {Miley}, {Nanni}, {Shimwell}, {Smith}, {Venemans}, and {Wagenveld}]{Gloudemans_2022}
A.~J. {Gloudemans} et al.
\newblock \emph{\aap}, 668:\penalty0 A27, Dec. 2022.
\newblock \doi{10.1051/0004-6361/202244763}.

\bibitem[{Gloudemans} et~al.(2023){Gloudemans}, {Saxena}, {Intema}, {Callingham}, {Duncan}, {R{\"o}ttgering}, {Belladitta}, {Hardcastle}, {Harikane}, and {Spingola}]{Gloudemans_2023}
A.~J. {Gloudemans} et al.
\newblock \emph{\aap}, 678:\penalty0 A128, Oct. 2023.
\newblock \doi{10.1051/0004-6361/202347582}.

\bibitem[{Gloudemans} et~al.(2025){Gloudemans}, {Duncan}, {Eilers}, {Farina}, {Harikane}, {Inayoshi}, {Lambrides}, and {Vardoulaki}]{Gloudemans_2025}
A.~J. {Gloudemans} et al.
\newblock \emph{\apj}, 986\penalty0 (2):\penalty0 130, June 2025.
\newblock \doi{10.3847/1538-4357/adddb9}.

\bibitem[{Goulding} et~al.(2023){Goulding}, {Greene}, {Setton}, {Labbe}, {Bezanson}, {Miller}, {Atek}, {Bogd{\'a}n}, {Brammer}, {Chemerynska}, {Cutler}, {Dayal}, {Fudamoto}, {Fujimoto}, {Furtak}, {Kokorev}, {Khullar}, {Leja}, {Marchesini}, {Natarajan}, {Nelson}, {Oesch}, {Pan}, {Papovich}, {Price}, {van Dokkum}, {Wang}, {Weaver}, {Whitaker}, and {Zitrin}]{Goulding_2023}
A.~D. {Goulding} et al.
\newblock \emph{\apjl}, 955\penalty0 (1):\penalty0 L24, Sept. 2023.
\newblock \doi{10.3847/2041-8213/acf7c5}.

\bibitem[{Greig} et~al.(2024){Greig}, {Mesinger}, {Ba{\~n}ados}, {Becker}, {Bosman}, {Chen}, {Davies}, {D'Odorico}, {Eilers}, {Gallerani}, {Haehnelt}, {Keating}, {Lai}, {Qin}, {Ryan-Weber}, {Satyavolu}, {Wang}, {Yang}, and {Zhu}]{Greig_2024}
B.~{Greig} et al.
\newblock \emph{\mnras}, 530\penalty0 (3):\penalty0 3208--3227, May 2024.
\newblock \doi{10.1093/mnras/stae1080}.

\bibitem[{Gupta} et~al.(2025){Gupta}, {Beniamini}, {Kumar}, and {Finkelstein}]{Gupta_2025}
O.~{Gupta}, P.~{Beniamini}, P.~{Kumar}, and S.~L. {Finkelstein}.
\newblock \emph{\apj}, 986\penalty0 (1):\penalty0 100, June 2025.
\newblock \doi{10.3847/1538-4357/add14c}.

\bibitem[{Harikane} et~al.(2023){Harikane}, {Zhang}, {Nakajima}, {Ouchi}, {Isobe}, {Ono}, {Hatano}, {Xu}, and {Umeda}]{Harikane_2023}
Y.~{Harikane} et al.
\newblock \emph{\apj}, 959\penalty0 (1):\penalty0 39, Dec. 2023.
\newblock \doi{10.3847/1538-4357/ad029e}.

\bibitem[{Hashimoto} et~al.(2020){Hashimoto}, {Goto}, {On}, {Lu}, {Santos}, {Ho}, {Wang}, {Kim}, and {Hsiao}]{Hashimoto_2020}
T.~{Hashimoto} et al.
\newblock \emph{\mnras}, 497\penalty0 (4):\penalty0 4107--4116, Oct. 2020.
\newblock \doi{10.1093/mnras/staa2238}.

\bibitem[{Heimersheim} et~al.(2022){Heimersheim}, {Sartorio}, {Fialkov}, and {Lorimer}]{Heimersheim_2022}
S.~{Heimersheim}, N.~S. {Sartorio}, A.~{Fialkov}, and D.~R. {Lorimer}.
\newblock \emph{\apj}, 933\penalty0 (1):\penalty0 57, July 2022.
\newblock \doi{10.3847/1538-4357/ac70c9}.

\bibitem[Hills et~al.(2018)Hills, Kulkarni, Meerburg, and Puchwein]{Hills_2018}
R.~Hills, G.~Kulkarni, P.~D. Meerburg, and E.~Puchwein.
\newblock \emph{Nature}, 564\penalty0 (7736):\penalty0 E32--E34, 2018.
\newblock \doi{10.1038/s41586-018-0796-5}.

\bibitem[{Ighina} et~al.(2021){Ighina}, {Belladitta}, {Caccianiga}, {Broderick}, {Drouart}, {Moretti}, and {Seymour}]{Ighina_2021}
L.~{Ighina} et al.
\newblock \emph{\aap}, 647:\penalty0 L11, Mar. 2021.
\newblock \doi{10.1051/0004-6361/202140362}.

\bibitem[{Ighina} et~al.(2023){Ighina}, {Caccianiga}, {Moretti}, {Belladitta}, {Broderick}, {Drouart}, {Leung}, and {Seymour}]{Ighina_2023}
L.~{Ighina} et al.
\newblock \emph{\mnras}, 519\penalty0 (2):\penalty0 2060--2068, Feb. 2023.
\newblock \doi{10.1093/mnras/stac3668}.

\bibitem[{Ighina} et~al.(2024){Ighina}, {Caccianiga}, {Moretti}, {Broderick}, {Leung}, {L{\'o}pez-S{\'a}nchez}, {Rigamonti}, {Seymour}, {An}, {Belladitta}, {Bisogni}, {Della Ceca}, {Drouart}, {Gargiulo}, and {Liu}]{Ighina_2024}
L.~{Ighina} et al.
\newblock \emph{\aap}, 692:\penalty0 A241, Dec. 2024.
\newblock \doi{10.1051/0004-6361/202451376}.

\bibitem[{Ighina} et~al.(2025){Ighina}, {Caccianiga}, {Moretti}, {Broderick}, {Leung}, {Rigamonti}, {Seymour}, {Afonso}, {Connor}, {Vignali}, {Wang}, {An}, {Arsioli}, {Bisogni}, {Dallacasa}, {Della Ceca}, {Liu}, {L{\'o}pez-S{\'a}nchez}, {Matute}, {Reynolds}, {Rossi}, {Spingola}, {Severgnini}, and {Tavecchio}]{Ighina_2025}
L.~{Ighina} et al.
\newblock \emph{\aap}, 698:\penalty0 A158, June 2025.
\newblock \doi{10.1051/0004-6361/202453650}.

\bibitem[{Ioka} and {M{\'e}sz{\'a}ros}(2005)]{Ioka_2005}
K.~{Ioka} and P.~{M{\'e}sz{\'a}ros}.
\newblock \emph{\apj}, 619\penalty0 (2):\penalty0 684--696, Feb. 2005.
\newblock \doi{10.1086/426785}.

\bibitem[{Jeon} et~al.(2025){Jeon}, {Bromm}, {Liu}, and {Finkelstein}]{Jeon_2025}
J.~{Jeon}, V.~{Bromm}, B.~{Liu}, and S.~L. {Finkelstein}.
\newblock \emph{\apj}, 979\penalty0 (2):\penalty0 127, Feb. 2025.
\newblock \doi{10.3847/1538-4357/ad9f3a}.

\bibitem[{Jiang} et~al.(2009){Jiang}, {Fan}, {Bian}, {Annis}, {Chiu}, {Jester}, {Lin}, {Lupton}, {Richards}, {Strauss}, {Malanushenko}, {Malanushenko}, and {Schneider}]{Jiang_2009}
L.~{Jiang} et al.
\newblock \emph{\aj}, 138\penalty0 (1):\penalty0 305--311, July 2009.
\newblock \doi{10.1088/0004-6256/138/1/305}.

\bibitem[{Kadota} et~al.(2023){Kadota}, {Villanueva-Domingo}, {Ichiki}, {Hasegawa}, and {Naruse}]{Kadota_2023}
K.~{Kadota} et al.
\newblock \emph{\jcap}, 2023\penalty0 (3):\penalty0 017, Mar. 2023.
\newblock \doi{10.1088/1475-7516/2023/03/017}.

\bibitem[{Kakiichi} et~al.(2025){Kakiichi}, {Jin}, {Wang}, {Meyer}, {Garaldi}, {Bosman}, {Davies}, {Fan}, {Trebitsch}, {Yang}, {Ba{\~n}ados}, {Champagne}, {Eilers}, {Hennawi}, {Sun}, {Wu}, {Zou}, {Kannan}, {Smith}, {Becker}, {D'Odorico}, {Connor}, {Liu}, {Protu{\v{s}}ov{\'a}}, {Walter}, and {Zhang}]{Kakiichi_2025}
K.~{Kakiichi} et al.
\newblock \emph{arXiv e-prints}, art. arXiv:2503.07074, Mar. 2025.
\newblock \doi{10.48550/arXiv.2503.07074}.

\bibitem[{Kashino} et~al.(2020){Kashino}, {Lilly}, {Shibuya}, {Ouchi}, and {Kashikawa}]{Kashino_2020}
D.~{Kashino} et al.
\newblock \emph{\apj}, 888\penalty0 (1):\penalty0 6, Jan. 2020.
\newblock \doi{10.3847/1538-4357/ab5a7d}.

\bibitem[{Kawasaki} et~al.(2021){Kawasaki}, {Nakano}, {Nakatsuka}, and {Sonomoto}]{Kawasaki_2021}
M.~{Kawasaki}, W.~{Nakano}, H.~{Nakatsuka}, and E.~{Sonomoto}.
\newblock \emph{\jcap}, 2021\penalty0 (4):\penalty0 019, Apr. 2021.
\newblock \doi{10.1088/1475-7516/2021/04/019}.

\bibitem[{Keating} et~al.(2020){Keating}, {Weinberger}, {Kulkarni}, {Haehnelt}, {Chardin}, and {Aubert}]{Keating_2020}
L.~C. {Keating} et al.
\newblock \emph{\mnras}, 491\penalty0 (2):\penalty0 1736--1745, Jan. 2020.
\newblock \doi{10.1093/mnras/stz3083}.

\bibitem[{Kulkarni} et~al.(2019){Kulkarni}, {Keating}, {Haehnelt}, {Bosman}, {Puchwein}, {Chardin}, and {Aubert}]{Kulkarni_2019}
G.~{Kulkarni} et al.
\newblock \emph{\mnras}, 485\penalty0 (1):\penalty0 L24--L28, May 2019.
\newblock \doi{10.1093/mnrasl/slz025}.

\bibitem[{Lidz} et~al.(2007){Lidz}, {McQuinn}, {Zaldarriaga}, {Hernquist}, and {Dutta}]{Lidz_2007}
A.~{Lidz} et al.
\newblock \emph{\apj}, 670\penalty0 (1):\penalty0 39--59, Nov. 2007.
\newblock \doi{10.1086/521974}.

\bibitem[{Liu} et~al.(2021){Liu}, {Wang}, {Momjian}, {Ba{\~n}ados}, {Zeimann}, {Willott}, {Matsuoka}, {Omont}, {Shao}, {Li}, and {Li}]{Liu_2021}
Y.~{Liu} et al.
\newblock \emph{\apj}, 908\penalty0 (2):\penalty0 124, Feb. 2021.
\newblock \doi{10.3847/1538-4357/abd3a8}.

\bibitem[{Mack} and {Wyithe}(2012)]{Mack_2012}
K.~J. {Mack} and J.~S.~B. {Wyithe}.
\newblock \emph{\mnras}, 425\penalty0 (4):\penalty0 2988--3001, Oct. 2012.
\newblock \doi{10.1111/j.1365-2966.2012.21561.x}.

\bibitem[{Maiolino} et~al.(2024){Maiolino}, {Scholtz}, {Witstok}, {Carniani}, {D'Eugenio}, {de Graaff}, {{\"U}bler}, {Tacchella}, {Curtis-Lake}, {Arribas}, {Bunker}, {Charlot}, {Chevallard}, {Curti}, {Looser}, {Maseda}, {Rawle}, {Rodr{\'\i}guez del Pino}, {Willott}, {Egami}, {Eisenstein}, {Hainline}, {Robertson}, {Williams}, {Willmer}, {Baker}, {Boyett}, {DeCoursey}, {Fabian}, {Helton}, {Ji}, {Jones}, {Kumari}, {Laporte}, {Nelson}, {Perna}, {Sandles}, {Shivaei}, and {Sun}]{Maiolino_2024}
R.~{Maiolino} et al.
\newblock \emph{\nat}, 627\penalty0 (8002):\penalty0 59--63, Mar. 2024.
\newblock \doi{10.1038/s41586-024-07052-5}.

\bibitem[{Maiolino} et~al.(2025){Maiolino}, {Risaliti}, {Signorini}, {Trefoloni}, {Juod{\v{z}}balis}, {Scholtz}, {{\"U}bler}, {D'Eugenio}, {Carniani}, {Fabian}, {Ji}, {Mazzolari}, {Bertola}, {Brusa}, {Bunker}, {Charlot}, {Comastri}, {Cresci}, {DeCoursey}, {Egami}, {Fiore}, {Gilli}, {Perna}, {Tacchella}, and {Venturi}]{Maiolino_2025}
R.~{Maiolino} et al.
\newblock \emph{\mnras}, 538\penalty0 (3):\penalty0 1921--1943, Apr. 2025.
\newblock \doi{10.1093/mnras/staf359}.

\bibitem[{Maity}(2024)]{Maity_2024}
B.~{Maity}.
\newblock \emph{\aap}, 689:\penalty0 A340, Sept. 2024.
\newblock \doi{10.1051/0004-6361/202451160}.

\bibitem[{Mazzolari} et~al.(2024){Mazzolari}, {Gilli}, {Maiolino}, {Prandoni}, {Delvecchio}, {Norman}, {Jimenez-Andrade}, {Belladitta}, {Vito}, {Momjian}, {Chiaberge}, {Trefoloni}, {Signorini}, {Ji}, {D'Amato}, {Risaliti}, {Baldi}, {Fabian}, {{\"U}bler}, {D'Eugenio}, {Scholtz}, {Juod{\v{z}}balis}, {Mignoli}, {Brusa}, {Murphy}, and {Muxlow}]{Mazzolari_2024}
G.~{Mazzolari} et al.
\newblock \emph{arXiv e-prints}, art. arXiv:2412.04224, Dec. 2024.
\newblock \doi{10.48550/arXiv.2412.04224}.

\bibitem[{McGreer} et~al.(2006){McGreer}, {Becker}, {Helfand}, and {White}]{McGreer_2006}
I.~D. {McGreer}, R.~H. {Becker}, D.~J. {Helfand}, and R.~L. {White}.
\newblock \emph{\apj}, 652\penalty0 (1):\penalty0 157--162, Nov. 2006.
\newblock \doi{10.1086/507767}.

\bibitem[McQuinn(2016)]{McQuinn_2016}
M.~McQuinn.
\newblock \emph{Annual Review of Astronomy and Astrophysics}, 54:\penalty0 313--362, 2016.
\newblock \doi{10.1146/annurev-astro-082214-122355}.

\bibitem[{Meiksin}(2011)]{Meiksin_2011}
A.~{Meiksin}.
\newblock \emph{\mnras}, 417\penalty0 (2):\penalty0 1480--1509, Oct. 2011.
\newblock \doi{10.1111/j.1365-2966.2011.19362.x}.

\bibitem[{Meiksin}(2020)]{Meiksin_2020}
A.~{Meiksin}.
\newblock \emph{\mnras}, 491\penalty0 (4):\penalty0 4884--4893, Feb. 2020.
\newblock \doi{10.1093/mnras/stz3395}.

\bibitem[Mertens et~al.(2020)]{Mertens_2020}
F.~G. Mertens et~al.
\newblock \emph{Monthly Notices of the Royal Astronomical Society}, 493\penalty0 (2):\penalty0 1662--1685, 2020.
\newblock \doi{10.1093/mnras/staa327}.

\bibitem[{Mesinger}(2010)]{Mesinger_2010}
A.~{Mesinger}.
\newblock \emph{\mnras}, 407\penalty0 (2):\penalty0 1328--1337, Sept. 2010.
\newblock \doi{10.1111/j.1365-2966.2010.16995.x}.

\bibitem[{Mesinger} et~al.(2011){Mesinger}, {Furlanetto}, and {Cen}]{Mesinger_2011}
A.~{Mesinger}, S.~{Furlanetto}, and R.~{Cen}.
\newblock \emph{\mnras}, 411\penalty0 (2):\penalty0 955--972, Feb. 2011.
\newblock \doi{10.1111/j.1365-2966.2010.17731.x}.

\bibitem[{Mirocha} and {Furlanetto}(2019)]{mirocha_2019}
J.~{Mirocha} and S.~R. {Furlanetto}.
\newblock \emph{\mnras}, 483\penalty0 (2):\penalty0 1980--1992, Feb. 2019.
\newblock \doi{10.1093/mnras/sty3260}.

\bibitem[Mu{\~n}oz and Loeb(2018)]{Munoz_2018}
J.~B. Mu{\~n}oz and A.~Loeb.
\newblock \emph{Nature}, 557\penalty0 (7707):\penalty0 684, 2018.
\newblock \doi{10.1038/s41586-018-0151-x}.

\bibitem[Mu{\~n}oz et~al.(2022)Mu{\~n}oz, Qin, Mesinger, Murray, Greig, and Mason]{Munoz_2021}
J.~B. Mu{\~n}oz et al.
\newblock \emph{Mon. Not. Roy. Astron. Soc.}, 511\penalty0 (3):\penalty0 3657--3681, 2022.
\newblock \doi{10.1093/mnras/stac185}.

\bibitem[{Murray} et~al.(2020){Murray}, {Greig}, {Mesinger}, {Mu{\~n}oz}, {Qin}, {Park}, and {Watkinson}]{Murray_2020}
S.~{Murray} et al.
\newblock \emph{The Journal of Open Source Software}, 5\penalty0 (54):\penalty0 2582, Oct. 2020.
\newblock \doi{10.21105/joss.02582}.

\bibitem[{Nakane} et~al.(2024){Nakane}, {Ouchi}, {Nakajima}, {Harikane}, {Ono}, {Umeda}, {Isobe}, {Zhang}, and {Xu}]{Nakane_2024}
M.~{Nakane} et al.
\newblock \emph{\apj}, 967\penalty0 (1):\penalty0 28, May 2024.
\newblock \doi{10.3847/1538-4357/ad38c2}.

\bibitem[{Nakatani} et~al.(2020){Nakatani}, {Fialkov}, and {Yoshida}]{Nakatani_2020}
R.~{Nakatani}, A.~{Fialkov}, and N.~{Yoshida}.
\newblock \emph{\apj}, 905\penalty0 (2):\penalty0 151, Dec. 2020.
\newblock \doi{10.3847/1538-4357/abc5b4}.

\bibitem[{Naruse} et~al.(2024){Naruse}, {Hasegawa}, {Kadota}, {Tashiro}, and {Ichiki}]{Naruse_2024}
G.~{Naruse} et al.
\newblock \emph{\jcap}, 2024\penalty0 (10):\penalty0 091, Oct. 2024.
\newblock \doi{10.1088/1475-7516/2024/10/091}.

\bibitem[{Nasir} and {D'Aloisio}(2020)]{Nasir_2020}
F.~{Nasir} and A.~{D'Aloisio}.
\newblock \emph{\mnras}, 494\penalty0 (3):\penalty0 3080--3094, Apr. 2020.
\newblock \doi{10.1093/mnras/staa894}.

\bibitem[{Nishizawa} et~al.(2025){Nishizawa}, {Natwariya}, and {Kadota}]{Nishizawa_2025}
A.~J. {Nishizawa}, P.~K. {Natwariya}, and K.~{Kadota}.
\newblock \emph{\prd}, 111\penalty0 (8):\penalty0 083546, Apr. 2025.
\newblock \doi{10.1103/PhysRevD.111.083546}.

\bibitem[{Niu} et~al.(2025){Niu}, {Li}, {Xu}, {Guo}, and {Zhang}]{Niu_2025}
Q.~{Niu} et al.
\newblock \emph{\apj}, 978\penalty0 (2):\penalty0 145, Jan. 2025.
\newblock \doi{10.3847/1538-4357/ad9b97}.

\bibitem[Paciga et~al.(2013)Paciga, Albert, Bandura, Chang, Gupta, Hirata, Odegova, Pen, Peterson, Roy, Shaw, Sigurdson, and Voytek]{Paciga_2013}
G.~Paciga et al.
\newblock \emph{Monthly Notices of the Royal Astronomical Society}, 433\penalty0 (1):\penalty0 639--647, 2013.
\newblock \doi{10.1093/mnras/stt753}.

\bibitem[{Pagano} et~al.(2020){Pagano}, {Delouis}, {Mottet}, {Puget}, and {Vibert}]{Pagano_2020}
L.~{Pagano} et al.
\newblock \emph{\aap}, 635:\penalty0 A99, Mar. 2020.
\newblock \doi{10.1051/0004-6361/201936630}.

\bibitem[{Pagano} and {Fronenberg}(2021)]{Pagano_2021}
M.~{Pagano} and H.~{Fronenberg}.
\newblock \emph{\mnras}, 505\penalty0 (2):\penalty0 2195--2206, Aug. 2021.
\newblock \doi{10.1093/mnras/stab1438}.

\bibitem[{Park} et~al.(2016){Park}, {Shapiro}, {Choi}, {Yoshida}, {Hirano}, and {Ahn}]{Park_2016}
H.~{Park} et al.
\newblock \emph{\apj}, 831\penalty0 (1):\penalty0 86, Nov. 2016.
\newblock \doi{10.3847/0004-637X/831/1/86}.

\bibitem[Patil et~al.(2017)]{Patil_2017}
A.~H. Patil et~al.
\newblock \emph{The Astrophysical Journal}, 838\penalty0 (1):\penalty0 65, 2017.
\newblock \doi{10.3847/1538-4357/aa63e7}.

\bibitem[{Patil} et~al.(2026){Patil}, {{\v{S}}oltinsk{\'y}}, {Maitra}, and {Kulkarni}]{Patil_2026}
S.~K. {Patil}, T.~{{\v{S}}oltinsk{\'y}}, S.~{Maitra}, and G.~{Kulkarni}.
\newblock \emph{\mnras}, Feb. 2026.
\newblock \doi{10.1093/mnras/stag236}.

\bibitem[{Pawlik} et~al.(2017){Pawlik}, {Rahmati}, {Schaye}, {Jeon}, and {Dalla Vecchia}]{Pawlik_2017}
A.~H. {Pawlik} et al.
\newblock \emph{\mnras}, 466\penalty0 (1):\penalty0 960--973, Apr. 2017.
\newblock \doi{10.1093/mnras/stw2869}.

\bibitem[Prelogovi{\'c} and Mesinger(2023)]{Prelogovic_2023}
D.~Prelogovi{\'c} and A.~Mesinger.
\newblock \emph{Mon. Not. Roy. Astron. Soc.}, 524\penalty0 (3):\penalty0 4239--4255, 2023.
\newblock \doi{10.1093/mnras/stad2027}.

\bibitem[{Puchwein} et~al.(2023){Puchwein}, {Bolton}, {Keating}, {Molaro}, {Gaikwad}, {Kulkarni}, {Haehnelt}, {Ir{\v{s}}i{\v{c}}}, {{\v{S}}oltinsk{\'y}}, {Viel}, {Aubert}, {Becker}, and {Meiksin}]{Puchwein_2023}
E.~{Puchwein} et al.
\newblock \emph{\mnras}, 519\penalty0 (4):\penalty0 6162--6183, Mar. 2023.
\newblock \doi{10.1093/mnras/stac3761}.

\bibitem[{Qin} et~al.(2021){Qin}, {Mesinger}, {Bosman}, and {Viel}]{Qin_2021}
Y.~{Qin}, A.~{Mesinger}, S.~E.~I. {Bosman}, and M.~{Viel}.
\newblock \emph{\mnras}, 506\penalty0 (2):\penalty0 2390--2407, Sept. 2021.
\newblock \doi{10.1093/mnras/stab1833}.

\bibitem[{Reis} et~al.(2020){Reis}, {Fialkov}, and {Barkana}]{Reis_2020}
I.~{Reis}, A.~{Fialkov}, and R.~{Barkana}.
\newblock \emph{\mnras}, 499\penalty0 (4):\penalty0 5993--6008, Dec. 2020.
\newblock \doi{10.1093/mnras/staa3091}.

\bibitem[{Sawyer} et~al.(2025){Sawyer}, {Bolton}, {Becker}, {Conaboy}, {Haehnelt}, {Keating}, {Kulkarni}, and {Puchwein}]{Sawyer_2025}
F.~{Sawyer} et al.
\newblock \emph{\mnras}, 540\penalty0 (3):\penalty0 2238--2252, July 2025.
\newblock \doi{10.1093/mnras/staf858}.

\bibitem[Saxena et~al.(2023)Saxena, Cole, Gazagnes, Meerburg, Weniger, and Witte]{Saxena_2023}
A.~Saxena et al.
\newblock \emph{Mon. Not. Roy. Astron. Soc.}, 525\penalty0 (4):\penalty0 6097--6111, 2023.
\newblock \doi{10.1093/mnras/stad2659}.

\bibitem[{Semelin}(2016)]{Semelin_2016}
B.~{Semelin}.
\newblock \emph{\mnras}, 455\penalty0 (1):\penalty0 962--973, Jan. 2016.
\newblock \doi{10.1093/mnras/stv2312}.

\bibitem[{Shao} et~al.(2023){Shao}, {Xu}, {Wang}, {Yang}, {Li}, {Zhang}, and {Chen}]{Shao_2023}
Y.~{Shao} et al.
\newblock \emph{Nature Astronomy}, 7:\penalty0 1116--1126, Sept. 2023.
\newblock \doi{10.1038/s41550-023-02024-7}.

\bibitem[{Shao} et~al.(2025){Shao}, {Sun}, {Zhao}, and {Zhang}]{Shao_2024}
Y.~{Shao}, T.-Y. {Sun}, M.-L. {Zhao}, and X.~{Zhang}.
\newblock \emph{\prd}, 112\penalty0 (6):\penalty0 063513, Sept. 2025.
\newblock \doi{10.1103/vpd1-1kyj}.

\bibitem[{Shaw} et~al.(2024){Shaw}, {Ghara}, {Beniamini}, {Zaroubi}, and {Kumar}]{Shaw_2024}
A.~K. {Shaw} et al.
\newblock \emph{arXiv e-prints}, art. arXiv:2409.03255, Sept. 2024.
\newblock \doi{10.48550/arXiv.2409.03255}.

\bibitem[{Shimabukuro}(2026)]{2026PhRvD.113h3525S}
H.~{Shimabukuro}.
\newblock \emph{\prd}, 113\penalty0 (8):\penalty0 083525, Apr. 2026.
\newblock \doi{10.1103/bl2w-crry}.

\bibitem[{Shimabukuro} et~al.(2014){Shimabukuro}, {Ichiki}, {Inoue}, and {Yokoyama}]{Shimabukuro_2014}
H.~{Shimabukuro}, K.~{Ichiki}, S.~{Inoue}, and S.~{Yokoyama}.
\newblock \emph{\prd}, 90\penalty0 (8):\penalty0 083003, Oct. 2014.
\newblock \doi{10.1103/PhysRevD.90.083003}.

\bibitem[{Shimabukuro} et~al.(2020{\natexlab{a}}){Shimabukuro}, {Ichiki}, and {Kadota}]{Shimabukuro_2020a}
H.~{Shimabukuro}, K.~{Ichiki}, and K.~{Kadota}.
\newblock \emph{\prd}, 101\penalty0 (4):\penalty0 043516, Feb. 2020{\natexlab{a}}.
\newblock \doi{10.1103/PhysRevD.101.043516}.

\bibitem[{Shimabukuro} et~al.(2020{\natexlab{b}}){Shimabukuro}, {Ichiki}, and {Kadota}]{Shimabukuro_2020b}
H.~{Shimabukuro}, K.~{Ichiki}, and K.~{Kadota}.
\newblock \emph{\prd}, 102\penalty0 (2):\penalty0 023522, July 2020{\natexlab{b}}.
\newblock \doi{10.1103/PhysRevD.102.023522}.

\bibitem[{Shimabukuro} et~al.(2023){Shimabukuro}, {Ichiki}, and {Kadota}]{Shimabukuro_2023}
H.~{Shimabukuro}, K.~{Ichiki}, and K.~{Kadota}.
\newblock \emph{\prd}, 107\penalty0 (12):\penalty0 123520, June 2023.
\newblock \doi{10.1103/PhysRevD.107.123520}.

\bibitem[{Shimabukuro} et~al.(2025){Shimabukuro}, {Xu}, and {Shao}]{Shimabukuro_2025a}
H.~{Shimabukuro}, Y.~{Xu}, and Y.~{Shao}.
\newblock \emph{\prd}, 112\penalty0 (6):\penalty0 063557, Sept. 2025.
\newblock \doi{10.1103/nxr4-14gb}.

\bibitem[{Sikder} et~al.(2025){Sikder}, {Park}, {Barkana}, {Yoshida}, and {Fialkov}]{Sikder_2025}
S.~{Sikder} et al.
\newblock \emph{arXiv e-prints}, art. arXiv:2509.11175, Sept. 2025.
\newblock \doi{10.48550/arXiv.2509.11175}.

\bibitem[Sims and Pober(2020)]{Sims_2019}
P.~H. Sims and J.~C. Pober.
\newblock \emph{Mon. Not. Roy. Astron. Soc.}, 492\penalty0 (1):\penalty0 22--38, 2020.
\newblock \doi{10.1093/mnras/stz3388}.

\bibitem[{Spergel} et~al.(2015){Spergel}, {Gehrels}, {Baltay}, {Bennett}, {Breckinridge}, {Donahue}, {Dressler}, {Gaudi}, {Greene}, {Guyon}, {Hirata}, {Kalirai}, {Kasdin}, {Macintosh}, {Moos}, {Perlmutter}, {Postman}, {Rauscher}, {Rhodes}, {Wang}, {Weinberg}, {Benford}, {Hudson}, {Jeong}, {Mellier}, {Traub}, {Yamada}, {Capak}, {Colbert}, {Masters}, {Penny}, {Savransky}, {Stern}, {Zimmerman}, {Barry}, {Bartusek}, {Carpenter}, {Cheng}, {Content}, {Dekens}, {Demers}, {Grady}, {Jackson}, {Kuan}, {Kruk}, {Melton}, {Nemati}, {Parvin}, {Poberezhskiy}, {Peddie}, {Ruffa}, {Wallace}, {Whipple}, {Wollack}, and {Zhao}]{Spergel_2015}
D.~{Spergel} et al.
\newblock \emph{arXiv e-prints}, art. arXiv:1503.03757, Mar. 2015.
\newblock \doi{10.48550/arXiv.1503.03757}.

\bibitem[{Spina} et~al.(2024){Spina}, {Bosman}, {Davies}, {Gaikwad}, and {Zhu}]{Spina_2024}
B.~{Spina} et al.
\newblock \emph{\aap}, 688:\penalty0 L26, Aug. 2024.
\newblock \doi{10.1051/0004-6361/202450798}.

\bibitem[{Sun} et~al.(2025){Sun}, {Shao}, {Li}, {Xu}, {Wang}, and {Zhang}]{Sun_2024}
T.-Y. {Sun} et al.
\newblock \emph{Communications Physics}, 8\penalty0 (1):\penalty0 220, May 2025.
\newblock \doi{10.1038/s42005-025-02139-5}.

\bibitem[Sun et~al.(2025)Sun, Foster, Liu, Mu{\~n}oz, and Slatyer]{Sun_2023}
Y.~Sun et al.
\newblock \emph{Phys. Rev. D}, 111\penalty0 (4):\penalty0 043015, 2025.
\newblock \doi{10.1103/PhysRevD.111.043015}.

\bibitem[{Takeuchi} et~al.(2016){Takeuchi}, {Morokuma-Matsui}, {Iono}, {Hirashita}, {Tee}, {Wang}, and {Momose}]{Takeuchi_2016}
T.~T. {Takeuchi} et al.
\newblock \emph{arXiv e-prints}, art. arXiv:1603.01938, Mar. 2016.
\newblock \doi{10.48550/arXiv.1603.01938}.

\bibitem[{Takeuchi} et~al.(2024){Takeuchi}, {Yata}, {Egashira}, {Aoshima}, {Ishii}, {Cooray}, {Nakanishi}, {Kohno}, and {Kono}]{Takeuchi_2024}
T.~T. {Takeuchi} et al.
\newblock \emph{\apjs}, 271\penalty0 (2):\penalty0 44, Apr. 2024.
\newblock \doi{10.3847/1538-4365/ad2517}.

\bibitem[{Thyagarajan}(2020)]{Thyagarajan_2020}
N.~{Thyagarajan}.
\newblock \emph{\apj}, 899\penalty0 (1):\penalty0 16, Aug. 2020.
\newblock \doi{10.3847/1538-4357/ab9e6d}.

\bibitem[Vasiliev and Shchekinov(2013)]{Vasiliev_2013}
E.~O. Vasiliev and Y.~A. Shchekinov.
\newblock \emph{Astrophys. J.}, 777:\penalty0 8, 2013.
\newblock \doi{10.1088/0004-637X/777/1/8}.

\bibitem[{{\v{{D}}urov\v{{c}}{\'i}kov{\'a}}} et~al.(2024){{\v{{D}}urov\v{{c}}{\'i}kov{\'a}}}, {Eilers}, {Chen}, {Satyavolu}, {Kulkarni}, {Simcoe}, {Keating}, {Haehnelt}, and {Ba{\~n}ados}]{Durovcikova_2024}
D.~{{\v{{D}}urov\v{{c}}{\'i}kov{\'a}}} et al.
\newblock \emph{\apj}, 969\penalty0 (2):\penalty0 162, July 2024.
\newblock \doi{10.3847/1538-4357/ad4888}.

\bibitem[{Villanueva-Domingo} and {Ichiki}(2023)]{Villanueva-Domingo_2023}
P.~{Villanueva-Domingo} and K.~{Ichiki}.
\newblock \emph{\pasj}, 75:\penalty0 S33--S49, Feb. 2023.
\newblock \doi{10.1093/pasj/psab119}.

\bibitem[{{\v{S}}oltinsk{\'y}} et~al.(2021){{\v{S}}oltinsk{\'y}}, {Bolton}, {Hatch}, {Haehnelt}, {Keating}, {Kulkarni}, {Puchwein}, {Chardin}, and {Aubert}]{Soltinsky_2021}
T.~{{\v{S}}oltinsk{\'y}} et al.
\newblock \emph{\mnras}, 506\penalty0 (4):\penalty0 5818--5835, Oct. 2021.
\newblock \doi{10.1093/mnras/stab1830}.

\bibitem[{{\v{S}}oltinsk{\'y}} et~al.(2023){{\v{S}}oltinsk{\'y}}, {Bolton}, {Molaro}, {Hatch}, {Haehnelt}, {Keating}, {Kulkarni}, and {Puchwein}]{Soltinsky_2023}
T.~{{\v{S}}oltinsk{\'y}} et al.
\newblock \emph{\mnras}, 519\penalty0 (2):\penalty0 3027--3045, Feb. 2023.
\newblock \doi{10.1093/mnras/stac3710}.

\bibitem[{{\v{S}}oltinsk{\'y}} et~al.(2025){{\v{S}}oltinsk{\'y}}, {Kulkarni}, {Tendulkar}, and {Bolton}]{Soltinsky_2025}
T.~{{\v{S}}oltinsk{\'y}}, G.~{Kulkarni}, S.~P. {Tendulkar}, and J.~S. {Bolton}.
\newblock \emph{\mnras}, 537\penalty0 (1):\penalty0 364--378, Feb. 2025.
\newblock \doi{10.1093/mnras/staf026}.

\bibitem[{Wang} et~al.(2019){Wang}, {Yang}, {Fan}, {Wu}, {Yue}, {Li}, {Bian}, {Jiang}, {Ba{\~n}ados}, {Schindler}, {Findlay}, {Davies}, {Decarli}, {Farina}, {Green}, {Hennawi}, {Huang}, {Mazzuccheli}, {McGreer}, {Venemans}, {Walter}, {Dye}, {Lyke}, {Myers}, and {Nunez}]{Wang_2019}
F.~{Wang} et al.
\newblock \emph{\apj}, 884\penalty0 (1):\penalty0 30, Oct. 2019.
\newblock \doi{10.3847/1538-4357/ab2be5}.

\bibitem[Weinberg(2003)]{Weinberg_2003}
D.~H. Weinberg.
\newblock In \emph{The Emergence of Cosmic Structure}, volume 666 of \emph{AIP Conference Proceedings}, pages 157--169, 2003.
\newblock \doi{10.1063/1.1581786}.

\bibitem[{Weinberger} et~al.(2019){Weinberger}, {Haehnelt}, and {Kulkarni}]{Weinberger_2019}
L.~H. {Weinberger}, M.~G. {Haehnelt}, and G.~{Kulkarni}.
\newblock \emph{\mnras}, 485\penalty0 (1):\penalty0 1350--1366, May 2019.
\newblock \doi{10.1093/mnras/stz481}.

\bibitem[{Willott} et~al.(2010){Willott}, {Albert}, {Arzoumanian}, {Bergeron}, {Crampton}, {Delorme}, {Hutchings}, {Omont}, {Reyl{\'e}}, and {Schade}]{Willott_2010}
C.~J. {Willott} et al.
\newblock \emph{\aj}, 140\penalty0 (2):\penalty0 546--560, Aug. 2010.
\newblock \doi{10.1088/0004-6256/140/2/546}.

\bibitem[{Wolf} et~al.(2024){Wolf}, {Salvato}, {Belladitta}, {Arcodia}, {Ciroi}, {Di Mille}, {Sbarrato}, {Buchner}, {H{\"a}mmerich}, {Wilms}, {Collmar}, {Dwelly}, {Merloni}, {Urrutia}, and {Nandra}]{Wolf_2024}
J.~{Wolf} et al.
\newblock \emph{\aap}, 691:\penalty0 A30, Nov. 2024.
\newblock \doi{10.1051/0004-6361/202451035}.

\bibitem[{Wouthuysen}(1952)]{Wouthuysen_1952}
S.~A. {Wouthuysen}.
\newblock \emph{AJ}, 57:\penalty0 31--32, Jan. 1952.
\newblock \doi{10.1086/106661}.

\bibitem[{Xu} et~al.(2009){Xu}, {Chen}, {Fan}, {Trac}, and {Cen}]{Xu_2009}
Y.~{Xu} et al.
\newblock \emph{\apj}, 704\penalty0 (2):\penalty0 1396--1404, Oct. 2009.
\newblock \doi{10.1088/0004-637X/704/2/1396}.

\bibitem[{Xu} et~al.(2011){Xu}, {Ferrara}, and {Chen}]{Xu_2011}
Y.~{Xu}, A.~{Ferrara}, and X.~{Chen}.
\newblock \emph{\mnras}, 410\penalty0 (3):\penalty0 2025--2042, Jan. 2011.
\newblock \doi{10.1111/j.1365-2966.2010.17579.x}.

\bibitem[{Yang} et~al.(2023){Yang}, {Fan}, {Gupta}, {Myers}, {Palanque-Delabrouille}, {Wang}, {Y{\`e}che}, {Aguilar}, {Ahlen}, {Alexander}, {Brooks}, {Dawson}, {de la Macorra}, {Dey}, {Dhungana}, {Fanning}, {Font-Ribera}, {Gontcho}, {Guy}, {Honscheid}, {Juneau}, {Kisner}, {Kremin}, {Le Guillou}, {Levi}, {Magneville}, {Martini}, {Meisner}, {Miquel}, {Moustakas}, {Nie}, {Percival}, {Poppett}, {Prada}, {Schlafly}, {Tarl{\'e}}, {Vargas Magana}, {Weaver}, {Wechsler}, {Zhou}, {Zhou}, and {Zou}]{Yang_2023}
J.~{Yang} et al.
\newblock \emph{\apjs}, 269\penalty0 (1):\penalty0 27, Nov. 2023.
\newblock \doi{10.3847/1538-4365/acf99b}.

\bibitem[{Yoshiura} and {Takahashi}(2018)]{Yoshiura_2018}
S.~{Yoshiura} and K.~{Takahashi}.
\newblock \emph{\mnras}, 473\penalty0 (2):\penalty0 1570--1575, Jan. 2018.
\newblock \doi{10.1093/mnras/stx2462}.

\bibitem[{Yue} et~al.(2009){Yue}, {Ciardi}, {Scannapieco}, and {Chen}]{Yue_2009}
B.~{Yue}, B.~{Ciardi}, E.~{Scannapieco}, and X.~{Chen}.
\newblock \emph{\mnras}, 398\penalty0 (4):\penalty0 2122--2133, Oct. 2009.
\newblock \doi{10.1111/j.1365-2966.2009.15261.x}.

\bibitem[{Zeimann} et~al.(2011){Zeimann}, {White}, {Becker}, {Hodge}, {Stanford}, and {Richards}]{Zeimann_2011}
G.~R. {Zeimann} et al.
\newblock \emph{\apj}, 736\penalty0 (1):\penalty0 57, July 2011.
\newblock \doi{10.1088/0004-637X/736/1/57}.

\bibitem[Zhang et~al.(2025)Zhang, Cang, Gao, and Li]{Zhang_2025}
Z.-X. Zhang, J.~Cang, Y.~Gao, and H.~Li.
\newblock \emph{JCAP}, 07:\penalty0 027, 2025.
\newblock \doi{10.1088/1475-7516/2025/07/027}.

\bibitem[{Zheng} et~al.(2026){Zheng}, {Montero-Camacho}, {Cai}, and {Mao}]{Zheng_2026}
Y.~{Zheng}, P.~{Montero-Camacho}, Z.~{Cai}, and Y.~{Mao}.
\newblock \emph{\mnras}, 545\penalty0 (4):\penalty0 staf2014, Feb. 2026.
\newblock \doi{10.1093/mnras/staf2014}.

\bibitem[{Zhu} et~al.(2021){Zhu}, {Becker}, {Bosman}, {Keating}, {Christenson}, {Ba{\~n}ados}, {Bian}, {Davies}, {D'Odorico}, {Eilers}, {Fan}, {Haehnelt}, {Kulkarni}, {Pallottini}, {Qin}, {Wang}, and {Yang}]{Zhu_2021}
Y.~{Zhu} et al.
\newblock \emph{\apj}, 923\penalty0 (2):\penalty0 223, Dec. 2021.
\newblock \doi{10.3847/1538-4357/ac26c2}.

\bibitem[{Zhu} et~al.(2022){Zhu}, {Becker}, {Bosman}, {Keating}, {D'Odorico}, {Davies}, {Christenson}, {Ba{\~n}ados}, {Bian}, {Bischetti}, {Chen}, {Davies}, {Eilers}, {Fan}, {Gaikwad}, {Greig}, {Haehnelt}, {Kulkarni}, {Lai}, {Pallottini}, {Qin}, {Ryan-Weber}, {Walter}, {Wang}, and {Yang}]{Zhu_2022}
Y.~{Zhu} et al.
\newblock \emph{\apj}, 932\penalty0 (2):\penalty0 76, June 2022.
\newblock \doi{10.3847/1538-4357/ac6e60}.

\bibitem[{Zhu} et~al.(2023){Zhu}, {Becker}, {Christenson}, {D'Aloisio}, {Bosman}, {Bakx}, {D'Odorico}, {Bischetti}, {Cain}, {Davies}, {Davies}, {Eilers}, {Fan}, {Gaikwad}, {Haehnelt}, {Keating}, {Kulkarni}, {Lai}, {Ma}, {Mesinger}, {Qin}, {Satyavolu}, {Takeuchi}, {Umehata}, and {Yang}]{Zhu_2023}
Y.~{Zhu} et al.
\newblock \emph{\apj}, 955\penalty0 (2):\penalty0 115, Oct. 2023.
\newblock \doi{10.3847/1538-4357/aceef4}.

\bibitem[{Zhu} et~al.(2024){Zhu}, {Becker}, {Bosman}, {Cain}, {Keating}, {Nasir}, {D'Odorico}, {Ba{\~n}ados}, {Bian}, {Bischetti}, {Bolton}, {Chen}, {D'Aloisio}, {Davies}, {Davies}, {Eilers}, {Fan}, {Gaikwad}, {Greig}, {Haehnelt}, {Kulkarni}, {Lai}, {Puchwein}, {Qin}, {Ryan-Weber}, {Satyavolu}, {Spina}, {Walter}, {Wang}, {Wolfson}, and {Yang}]{Zhu_2024}
Y.~{Zhu} et al.
\newblock \emph{\mnras}, 533\penalty0 (1):\penalty0 L49--L56, Sept. 2024.
\newblock \doi{10.1093/mnrasl/slae061}.

\bibitem[{Ziegler} et~al.(2024){Ziegler}, {Shapiro}, {Dawoodbhoy}, {Beniamini}, {Kumar}, {Freese}, {Ocvirk}, {Aubert}, {Lewis}, {Teyssier}, {Park}, {Ahn}, {Sorce}, {Iliev}, {Yepes}, and {Gottlober}]{Ziegler_2025}
J.~J. {Ziegler} et al.
\newblock \emph{arXiv e-prints}, art. arXiv:2411.02699, Nov. 2024.
\newblock \doi{10.48550/arXiv.2411.02699}.

\end{thebibliography}

\end{document}